\documentclass[
 preprint,
 amsmath,amssymb,
 aps,pre
]{revtex4-2}
\usepackage[margin=1in]{geometry}
\usepackage{graphicx}
\usepackage{setspace}
\usepackage{xr}
\usepackage{caption}
\usepackage{todonotes}
\usepackage{caption}
\usepackage{subcaption}
\pdfoutput=1 

\DeclareMathOperator{\sech}{sech}

\begin{document} 
\title{Phase space analysis of nonlinear wave propagation in a bistable mechanical metamaterial with a defect}
\author{Mohammed A. Mohammed}
\email{mmohammed9@huskers.unl.edu}
\author{Piyush Grover}%
 \email{piyush.grover@unl.edu}
\affiliation{
 Mechanical and Materials Engineering, \\University of Nebraska-Lincoln, Lincoln, Nebraska, USA}
 \date{\today}
\begin{abstract}
    We study the dynamics of solitary waves traveling in a one-dimensional chain of bistable elements in the presence of a local inhomogeneity (`defect’). Numerical simulations reveal that depending upon its initial speed, an incoming solitary wave can get transmitted, captured or reflected upon interaction with the defect. The dynamics are dominated by energy exchange between the wave and a breather mode localized at the defect. We derive a reduced-order two degree of freedom Hamiltonian model for wave-breather interaction, and analyze it using dynamical systems techniques. Lobe dynamics analysis reveals the fine structure of phase space that leads to the complicated dynamics in this system. This work is a step towards developing a rational approach to defect engineering for manipulating nonlinear waves in mechanical metamaterials.
\end{abstract}
\maketitle

\section{Introduction}
Acoustic metamaterials \cite{bertoldi2017flexible} are (generally) periodic structures assembled using artificially engineered units, and designed to possess unconventional mechanical wave propagation characteristics. This class of mechanical metamaterials has potential applications in vibration control \cite{deng2020characterization}, energy harvesting \cite{hwang2022topological}, mechanical computing \cite{ion2017digital}, precision sensing \cite{xinjing2019acoustic} and cloaking \cite{cummer2016controlling}. The key to realizing the vast promise of such metamaterials lies in developing rational design and control techniques for manipulating the flow of energy in these systems. Since the dispersion relation contains all the information about the propagation and growth (or decay) of linear waves, the research in \emph{linear} acoustic metamaterials has focussed on developing techniques for tailoring the dispersion relation \cite{hussein2014dynamics,miniaci2021design}.

For \emph{nonlinear} metamaterials \cite{deng2021nonlinear}, the design space is vastly less explored due to the increased complexity of the nonlinear dynamics of wave propagation. One of the popular architectures consists of a one-dimensional chain of bistable elements connected by linear springs \cite{nadkarni2014dynamics,hwang2018input,hwang2018solitary}. This class of metamaterials supports the propagation of solitary waves \cite{dauxois2006physics,friesecke1994existence}, i.e., large amplitude, spatially localized waves that can travel large distances without distortion. Depending on the system geometry, continuum approximations of such systems (corresponding to inter-mass distance going to $0$) reduce to variants of the canonical nonlinear partial differential equations (PDEs) such as the Sine-Gordon and Klein-Gordon equations. This connection has been exploited in the analysis of 1D and 2D acoustic bistable metamaterial systems in previous works \cite{deng2018metamaterials,hwang2018input,hwang2018solitary,deng2020nonlinear}.
The propagation of solitary waves in such nonlinear structures can be tailored by introducing suitable spatial variations in mass or spring stiffness. In previous works, the effect of introducing inhomogeneity in bistable chains has been studied analytically in the weakly nonlinear regime for the case of 1D chain with spatially graded stiffness \cite{hwang2018solitary}, and numerically in the case of 1D or 2D structures with localized inhomogeneities (`defects') in mass and stiffness  \cite{hwang2018input,deng2018metamaterials}. The presence of a defect can give rise to an oscillatory mode (`breather') localized at the defect \cite{dauxois2006physics,hwang2018input,figotin1997localized}.

In this paper, we employ methods of dynamical systems theory to gain a deeper understanding of the dynamics of solitary waves in a 1D chain of bistable elements in the presence of a stiffness defect. Numerical simulations of the discrete chain show that depending on its speed, an incoming solitary wave can be transmitted, captured or reflected back upon interaction with the defect. To understand these numerical results, following earlier work on Sine-Gordon equation with a defect \cite{goodman2002interaction}, we derive a two degree of freedom (DOF) reduced-order model for the continuum approximation of this system using the method of collective coordinates. The two DOFs correspond to the position of the solitary wave, and the amplitude of the breather. This model is shown to capture some qualitative aspects of dynamics. The analysis of phase space transport in the system reveals the organizing structures that delineate the sets of initial conditions of solitary waves leading to qualitatively different outcomes after interaction with the defect.

\section{System Model and Coherent Structures}
\subsection{System without a defect}
\subsubsection{N-DOF System}
\begin{figure}
\centering
\includegraphics[width=.485\textwidth]{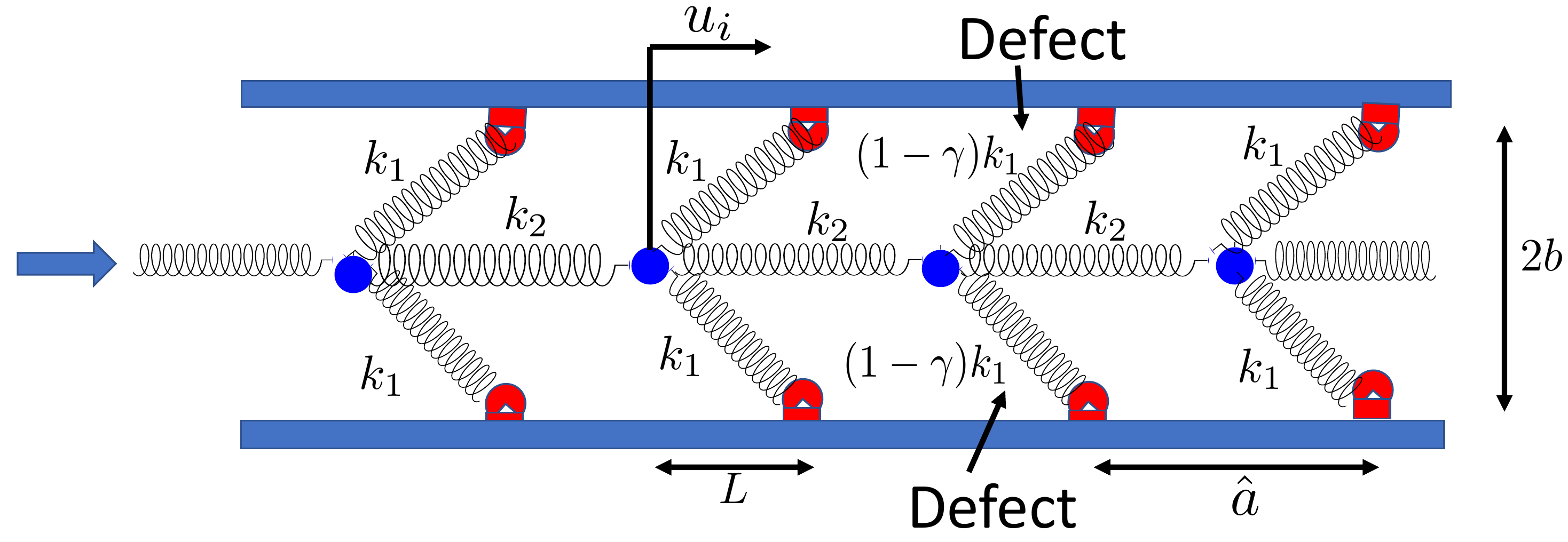}
\caption{\footnotesize An infinite chain of bistable elements with a defect in stiffness of a pair onsite springs. The distance between neighboring ground joints is $\hat{a}$. In the unstressed configuration, $L$ is the horizontal distance between the mass at site $i$ and the corresponding ground joints. The displacement of the mass at site $i$ is $u_i$. In the absence of defects, each onsite spring has stiffness $k_1$, and each intersite spring has stiffness $k_2$. The parameter $0\leq \gamma< 1$ controls the size of the defect. }
\label{fig:bistable_chain}
\end{figure}

We begin by discussing the defect-free chain of bistable elements studied in \cite{nadkarni2014dynamics}, and summarize their main results in the fully nonlinear (large amplitude) regime. The periodic chain consists of $N$ bistable units connected
by (`intersite') linear springs of stiffness $k_2$, see Fig. \ref{fig:bistable_chain}. Each unit consists of two (`onsite') identical
linear elastic springs with stiffness
$k_1$ and unstressed length $l_0$, connected to point a mass $m$ in a symmetric fashion. The other ends of the springs are fixed to the ground via joints that allow rotation. It was shown in \cite{nadkarni2014dynamics} that this system supports stable propagation of solitary waves (displacement `kinks') in the large amplitude limit. Let $u_i$ denote the displacement of the $i$th mass from the unstressed configuration. From Fig. \ref{fig:bistable_chain}, we get $l_0^2=L^2+b^2$. The length of each of the two onsite springs at site $i$ is $l(u_i)=\sqrt{(L-u_{i})^{2}+b^2}$, and the total force exerted by them on the $i$th mass is 	$F(u_{i})=2 k_{1}(L-u_{i})\dfrac{l(u_{i})-l_{0}}{l(u_{i})}$. Following \cite{nadkarni2014dynamics}, we non-dimensionalize the system using the relations: $\bar{u}_{i}=\dfrac{u}{L}$, $K_r=\dfrac{k_2}{k_1}$,$\:\bar{l}(\bar{u}_{i})=\sqrt{(1-\bar{u}_{i})^{2}+d^2}$, $d=\dfrac{b}{L}$, and $\bar{l}_{0}=\sqrt{1+d^2}$.

The dimensionless equations of motion of the chain are 
\begin{equation}\label{Dimensionless_EOM} 
	 \bar{u}_{i,\bar{t}\bar{t}}+K_{r}(-\bar{u}_{i+1}+2\bar{u}_{i}-\bar{u}_{i-1})-\bar{F}(\bar{u}_{i})=0,
\end{equation}
where 
$	\bar{F}(\bar{u}_{i})=-\dfrac{\partial\psi(\bar{u}_i)}{\partial \bar{u}_i}=\dfrac{F(\bar{u}_{i})}{k_{1}L}=2(1-\bar{u}_{i})(1-\dfrac{\bar{l}_{0}}{\bar{l}(\bar{u}_{i})})$. Finally, $\psi(\bar{u}_i)=\left [\sqrt{(1-\bar{u}_i)^2+d^2}-\sqrt{1+d^2}\right ]^2$ is the dimensionless nonlinear spring potential.

\subsubsection{Continuum limit}
Let the $i$th mass be initially located at $x_{i}=i\hat{a}$, and define non-dimensionalized quantities
\begin{equation}\label{Normlized Pos} 
	 \bar{x}_{i}=\frac{x_{i}}{\hat{a}},\;\;\; \Delta\bar{x}_{i}=\bar{x}_{i}-\bar{x}_{i-1}=1.
\end{equation}
In the continuum limit $\hat{a}\rightarrow 0$, Taylor expansion yields
\begin{equation}\label{TaySeru} 
	 \bar{u}_{i\pm1}=\bar{u}_{i}\pm \frac{\partial\bar{u}_{i}}{\partial\bar{x}}+\frac{1}{2} \frac{\partial^{2}\bar{u}_{i}}{\partial\bar{x}^{2}}+O(3).
\end{equation}
Substituting (\ref{TaySeru}) into (\ref{Dimensionless_EOM}) gives the nonlinear PDE
\begin{equation}\label{PDE} 
	 \bar{u}_{,\bar{t}\bar{t}}-K_{r} \bar{u}_{,\bar{x}\bar{x}}-\bar{F}(\bar{u})=0.
\end{equation}
For the rest of the paper, we drop the overbars for convenience. In the large amplitude limit, this system supports a solitary wave solution of the form $\tilde{u}_k(x,t)=\hat{u}_k(x-vt)=\hat{u}_k(z)$, where $v$ is the propagation velocity and $z=x-vt$ is a reduced variable. The solitary wave satisfies the implicit equation in $\hat{u}_k(z)$, 
\begin{align}\label{eq:exact_kink}
 \ln{\left[b_1(\hat{u}_k)+\frac{b_1(\hat{u}_k)}{b_2(\hat{u}_k)}\sqrt{1+d^2}\right]}+\frac{\sqrt{1+d^2}}{2}\ln{\left[\frac{1-b_1(\hat{u}_k)}{1+b_1(\hat{u}_k)}\frac{1-b_2(\hat{u}_k)}{1+b_2(\hat{u}_k)}\right]}=\sqrt{\frac{2}{C_0^2-v^2}}(z-z_0), 
\end{align}where 
$z_0$ is a constant of integration, $C_0=\sqrt{K_r}$ , $b_1(\hat{u}_k)=\hat{u}_k-1$, and $b_2(\hat{u}_k)=\dfrac{\sqrt{1+d^2}(\hat{u}_k-1)}{\sqrt{(\hat{u}_k-1)^2+d^2}}$.
\subsubsection{Dispersion Relation}
The linearization of (\ref{PDE}) about $u=0$ is 
\begin{equation}\label{LinearPDE} 
	u_{,tt}-K_{r} u_{,xx}+\omega_{0}^{2}u=0,
\end{equation}
where $\omega_0^2=\dfrac{2}{1+d^2}$. By looking for solutions of the form $u(x,t)=\tilde u e^{i(qx-\omega t)}$, we obtain the dispersion relation 
\begin{equation}\label{eq:dispersalb} 
	\omega_L=\sqrt{\omega_{0}^{2}+K_{r}q^{2}},
\end{equation} where $q \in\left (0,\infty\right)$ is the (spatial) wavenumber, and $\omega_L$ is the (temporal) frequency.
This relation shows that there is a bandgap in the system, i.e., the defectless system supports linear waves limited to the frequency range $\omega_L\in\left (w_0,\infty\right)$.

\subsection{System with a defect }

A localized inhomogeneity is now introduced in the bistable chain by modifying the onsite spring stiffness to be $(1-\gamma)k_1$ at the origin. Thus, the new nonlinear and linearized PDEs are
\begin{equation}\label{eq:PDE_def} 
	 {u}_{,tt}-K_{r} {u}_{,xx}-(1-\gamma \delta(x)){F}({u})=0,
\end{equation} and
\begin{equation}\label{eq:linPDE_def} 
	 u_{,tt}-K_{r} u_{,xx}+(1-\gamma\delta(x))\omega_0^{2}u=0,
\end{equation} respectively, 
where $0<\gamma<1$ is the defect magnitude, and $\delta(x)$ is the Dirac delta. Motivated by previous works \cite{fei1992resonantphi4,figotin1997localized,dauxois2006physics,kivshar1991resonant,goodman2002interaction,goodman2004strong,zhou2017kink}, we explore the possibility that this `small' perturbation of (\ref{LinearPDE}) can support spatially localized coherent structures with frequencies that lie in the bandgap $\left(0,\omega_0\right)$. 
Inserting the ansatz $u(x,t)=\phi(x)e^{i\omega t}$ into the linearized equation (\ref{eq:linPDE_def}) yields
\begin{equation}\label{eq32} 
	 K_{r} \phi_{,xx}+(\omega^{2}-\omega_{0}^{2})\phi=-\gamma\delta(x)\omega_{0}^{2}\phi.
\end{equation}
Let us first solve (\ref{eq32}) in the region $x\in \left(-\infty,-\xi\right)\cup\left(\xi,\infty\right)$, where  $0<\xi\ll 1$. In this region, (\ref{eq32}) reduces to
\begin{equation}\label{eq33} 
	 K_{r} \phi_{,xx}+(\omega^{2}-\omega_{0}^{2})\phi=0.
\end{equation}
Putting $\phi=C e^{\kappa x}$, 

\begin{equation}\label{eq34} 
	 K_{r} \kappa^{2} \phi_{,xx}+(\omega^{2}-\omega_{0}^{2})\phi=0,
\end{equation}
where
\begin{equation}\label{eq35} 
	 \kappa^{2}=\frac{\omega_{0}^{2}-\omega^{2}}{K_{r}}.
\end{equation}
The above equation implies that $\kappa$ will be real as long as $\omega$ lies in the bandgap, i.e,  $\omega < \omega_{0}$. In that case, (\ref{eq34}) has the solution of the form
\begin{equation}\label{eq36} 
	 \phi(x)=C_{1} e^{\kappa x}+C_{2} e^{-\kappa x}.
\end{equation}
Clearly, $\phi$ will blow up as $x\rightarrow-\infty$ unless $C_{2}$ is zero on $\left(-\infty,-\xi\right)$. Similarly, $C_{1}$ must be zero on $\left(\xi,\infty\right)$. Thus, the solution has the form:
\begin{equation}\label{eq37} 
	\phi(x)=\left\{
        \begin{array}{ll}
             C_{1} e^{\kappa x}\;\;\;\;, \quad x<0,  \\
            C_{2} e^{-\kappa x}\;\;, \quad x>0.
        \end{array}
    \right.
\end{equation}
Continuity of the solution at $x=0$ requires $C_{1}=C_{2}$, and hence, $\phi(x)=C_{1} e^{-\kappa |x|}$. Hence, the solution is localized in space, and periodic in time, i.e., a breather \cite{kivshar1991resonant}.

To compute the breather frequency $\omega$, we integrate both sides of (\ref{eq32}) over the interval $-\xi<x<\xi$, yielding
\begin{equation}\label{eq39} 
	\int_{-\xi}^{\xi}K_{r} \phi_{,xx} \,dx+\int_{-\xi}^{\xi}(\omega^{2}-\omega_{0}^{2})\phi \,dx=\int_{-\xi}^{\xi}-\gamma\delta(x)\omega_{0}^{2}\phi dx,
\end{equation}

\begin{equation}\label{eq40} 
	K_{r}(\phi_{,x}|_{x=\xi}-\phi_{,x}|_{x=-\xi})+\int_{-\xi}^{\xi}(\omega^{2}-\omega_{0}^{2})\phi \,dx=\int_{-\xi}^{\xi}-\gamma\delta(x)\omega_{0}^{2}\phi dx.
\end{equation}
Taking the limit $\xi\rightarrow0$, the second term of the left hand side of (\ref{eq40}) vanishes since $\phi$ is finite, and we obtain
\begin{equation}\label{eq41} 
	K_{r}(-\kappa\;C_{1}\;e^{-\kappa x}-\kappa\; C_{1}\; e^{-\kappa x})=-\gamma\omega_{0}^{2}\phi(0)=-\gamma\omega_{0}^{2}C_{1},
\end{equation}
and
\begin{equation}\label{kappa} 
	\kappa=\frac{\gamma\omega_{0}^{2}}{2K_{r}}.
\end{equation}
Substitution of the relation in (\ref{kappa}) into (\ref{eq35}) yields the relation

\begin{equation}\label{eq:omega} 
	\omega=\omega_{0}\sqrt{1-\frac{\gamma^{2}\omega_{0}^{2}}{4K_{r}}}.
\end{equation}
Finally, the full breather solution is 
\begin{equation}\label{eq47} 
	\hat{u}_b(x,t)=C_1\cos{(\omega t+\theta)}\phi(x)=C_1 \cos{(\omega t+\theta)}e^{-\dfrac{\gamma\omega_{0}^{2}}{2K_{r}}\;\left|x\right|},
\end{equation}
where $C_1$ and $\theta$ are constants that depend on the initial conditions.
We emphasize that the breather is an exact solution of the \emph{linearized} PDE (\ref{eq:linPDE_def}). This solution is expected to decay anomalously slowly in the nonlinear system (\ref{eq:PDE_def}) due to radiation damping effects, rendering it `metastable' \cite{goodman2002interaction,soffer1999resonances}.

 \subsubsection{Numerical simulations of the discrete and continuum models for the system with defect}

\begin{figure}
     \centering
     \begin{subfigure}[b]{0.45\textwidth}
         \centering
         \includegraphics[width=\textwidth]{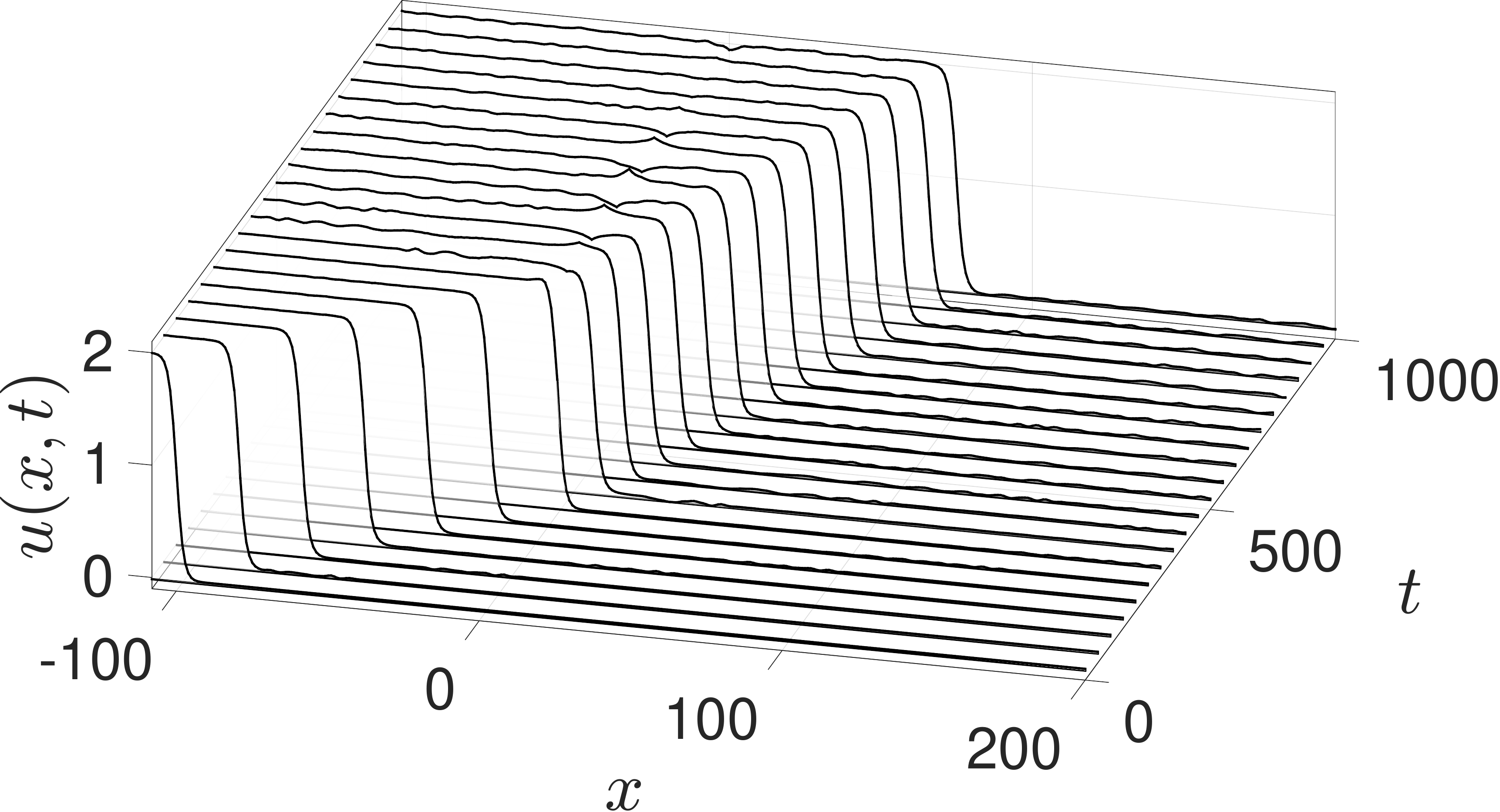}
         \caption{\footnotesize Discrete model}
     \end{subfigure}
     \hfill
     \begin{subfigure}[b]{0.45\textwidth}
         \centering
         \includegraphics[width=\textwidth]{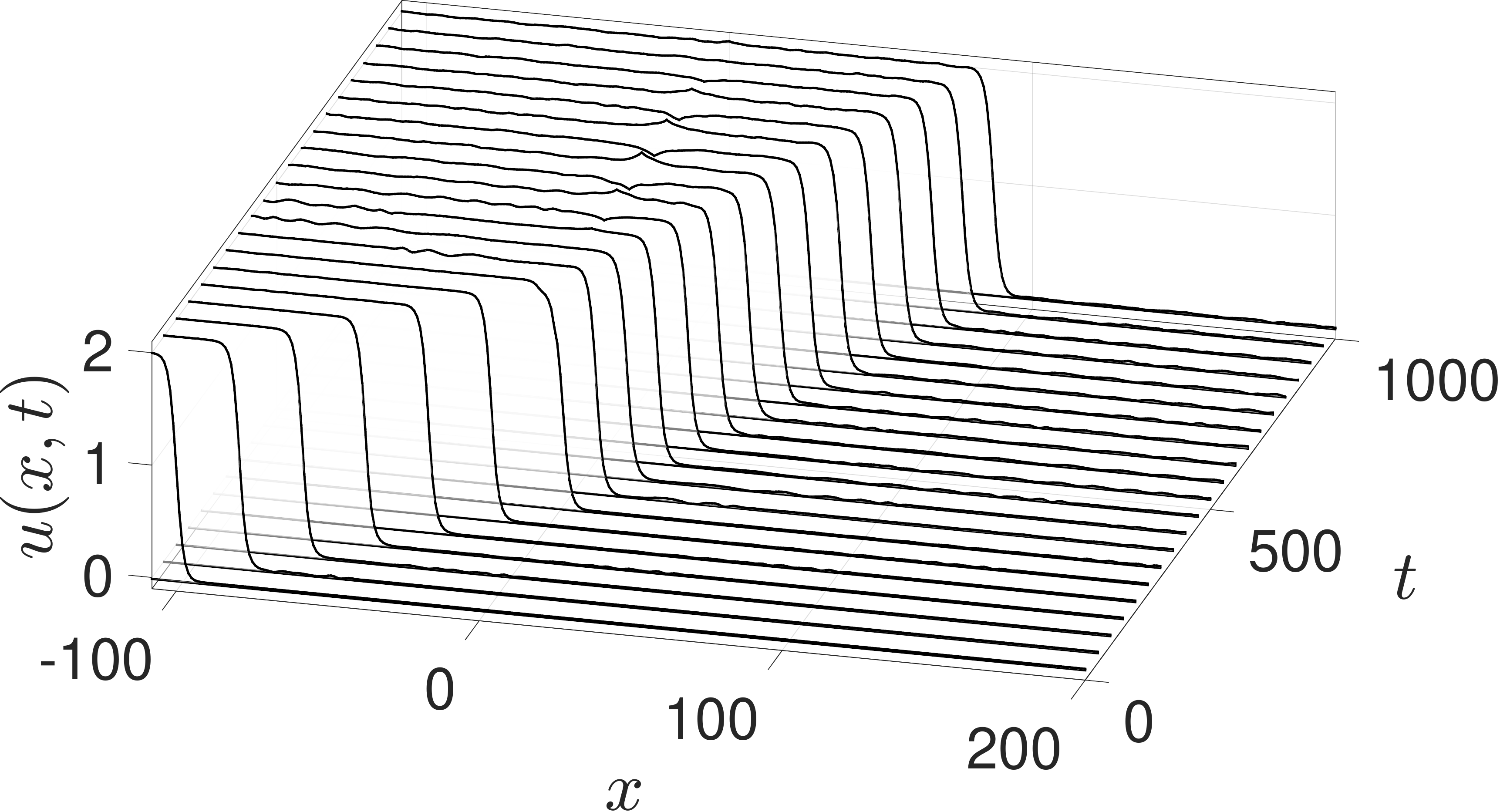}
         \caption{\footnotesize Continuum model}
     \end{subfigure}
        \caption{\footnotesize Space-time plot of an incoming wave transmitting across the defect in the (a) discrete and (b) continuum models. The breather mode gets `activated' once the wave has passed the defect. The parameters are $d=1,K_r=2, $ and $ \gamma=0.9$. }
        \label{fig:def_prop}
\end{figure}
\begin{figure}
     \centering
     \begin{subfigure}[b]{0.45\textwidth}
         \centering
         \includegraphics[width=\textwidth]{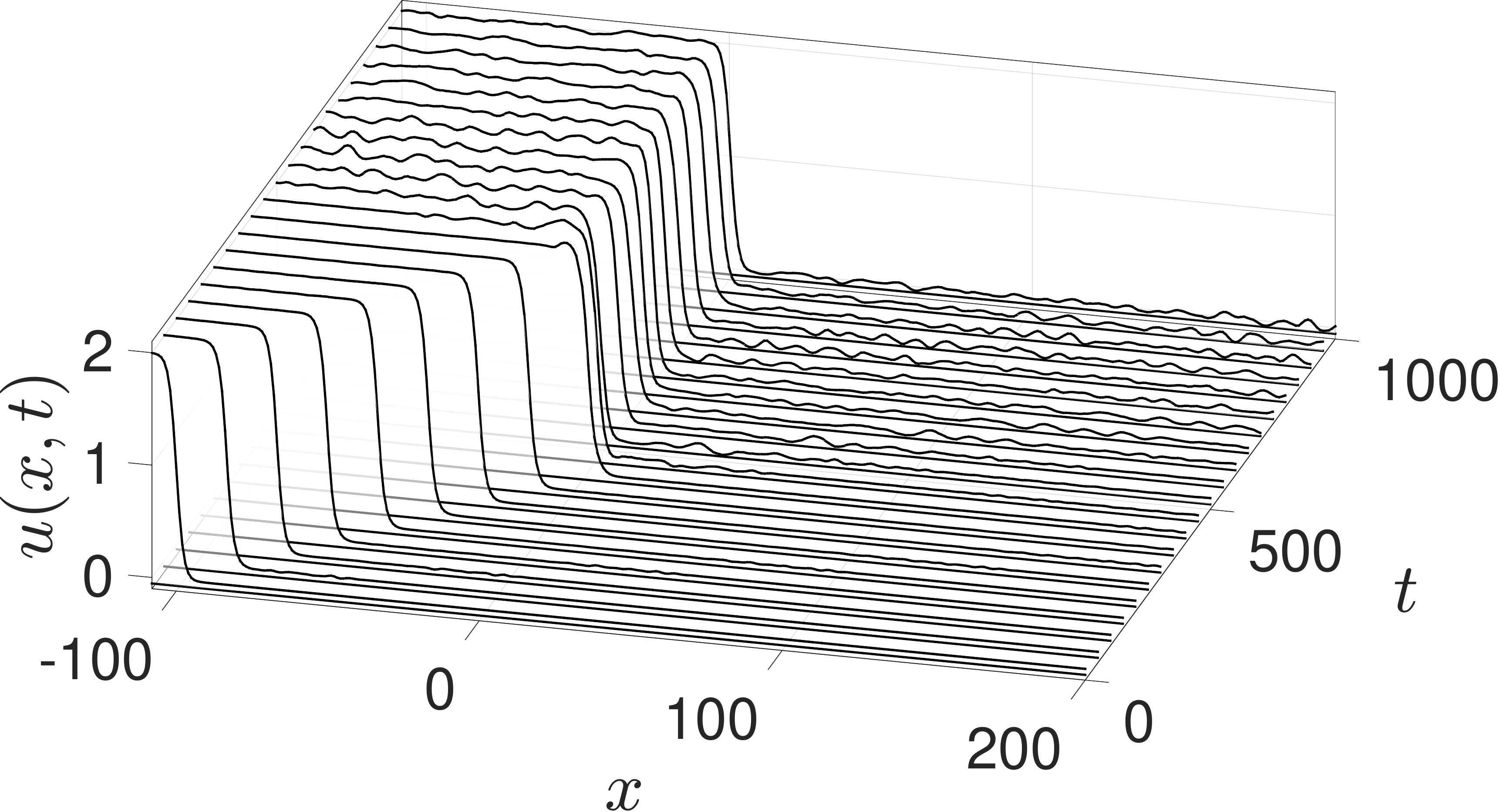}
         \caption{\footnotesize Discrete model}
     \end{subfigure}
     \hfill
     \begin{subfigure}[b]{0.45\textwidth}
         \centering
         \includegraphics[width=\textwidth]{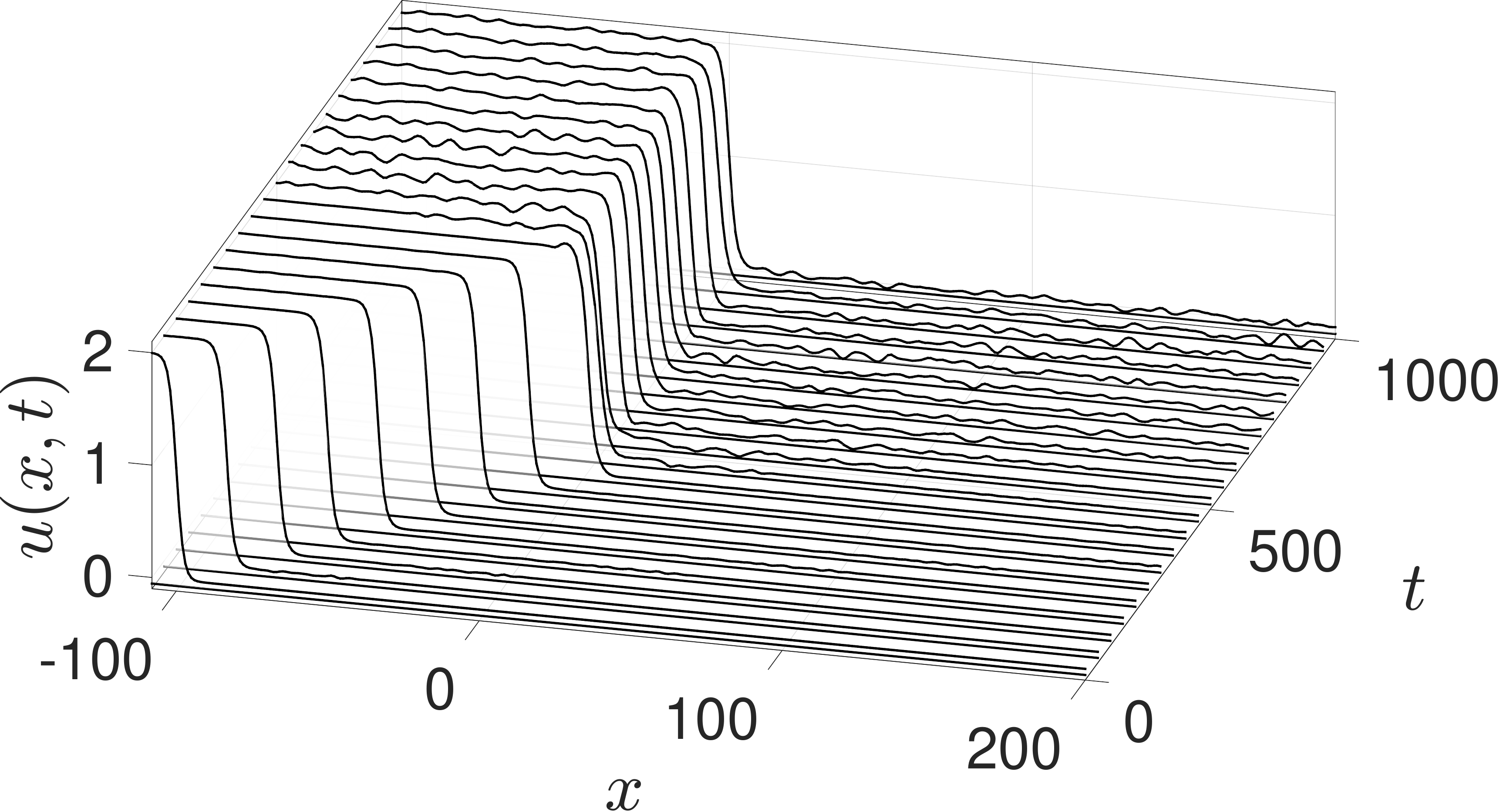}
         \caption{\footnotesize Continuum model}
     \end{subfigure}
        \caption{\footnotesize Space-time plot of an incoming wave getting captured at the defect in the (a) discrete and (b) continuum models. The parameters are $d=1,K_r=2, $ and $ \gamma=0.9$.}
        \label{fig:def_cap}
\end{figure}
\begin{figure}
     \centering
     \begin{subfigure}[b]{0.45\textwidth}
         \centering
         \includegraphics[width=\textwidth]{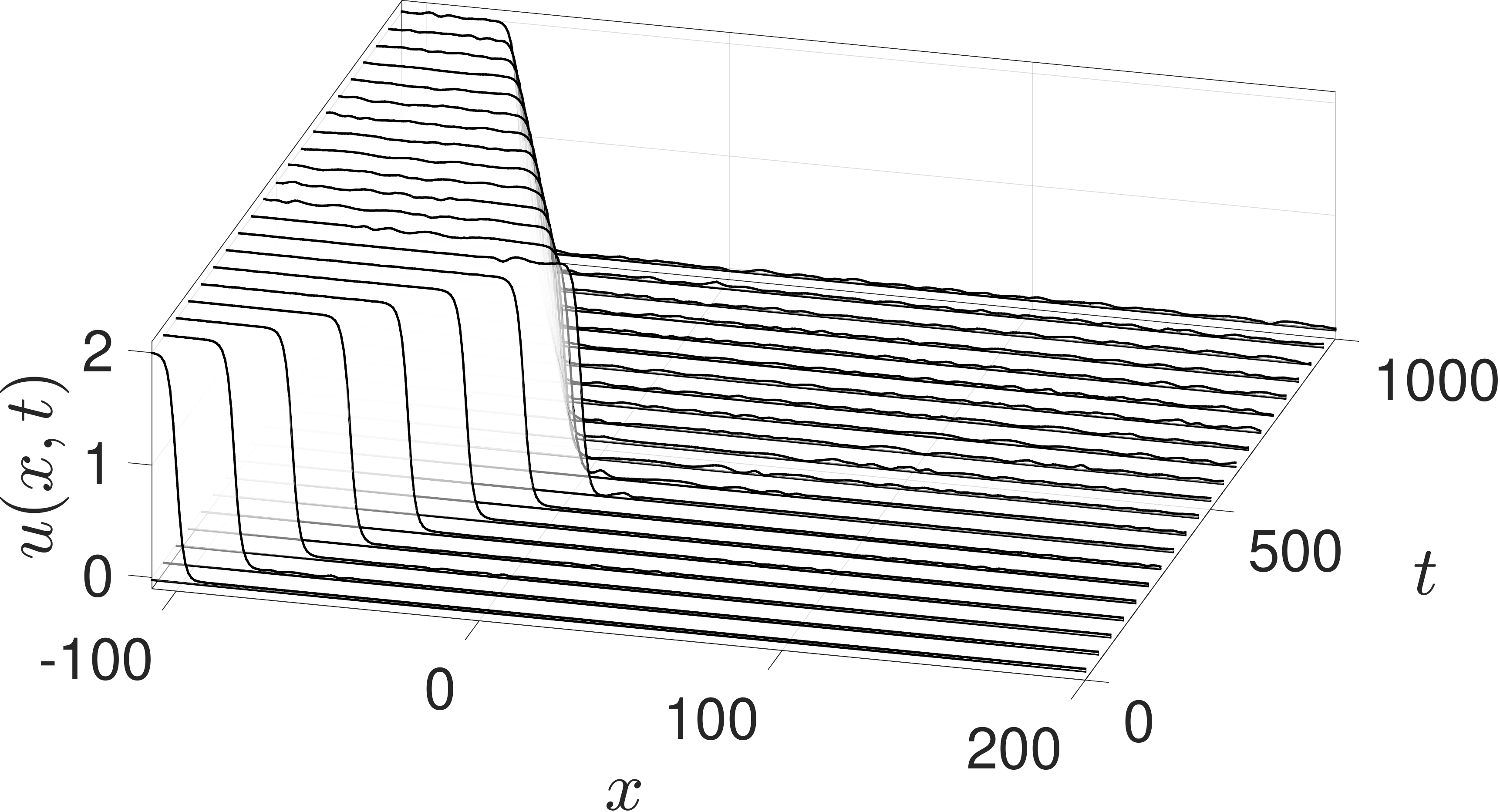}
         \caption{\footnotesize Discrete model}
     \end{subfigure}
     \hfill
     \begin{subfigure}[b]{0.45\textwidth}
         \centering
         \includegraphics[width=\textwidth]{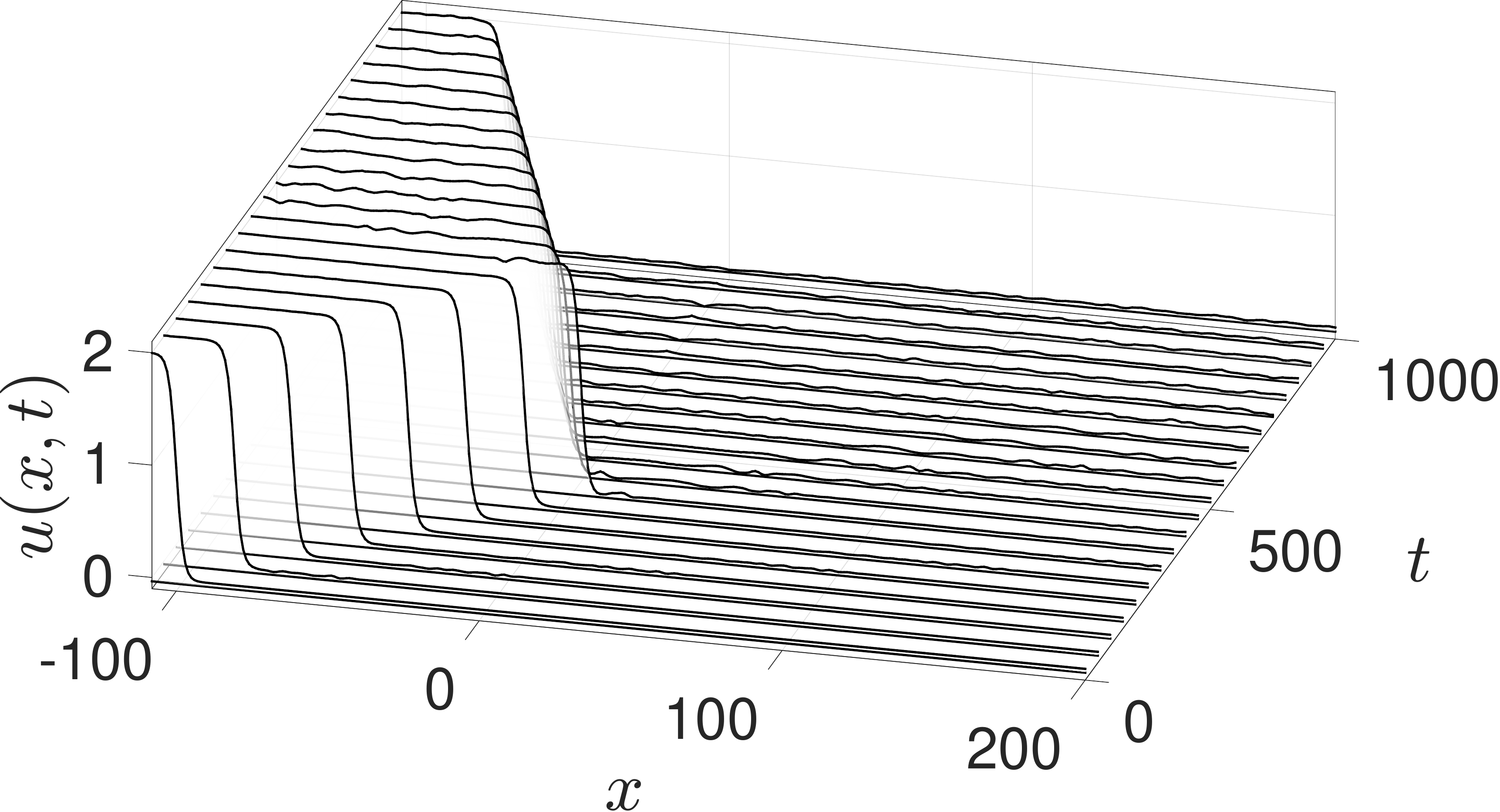}
         \caption{\footnotesize Continuum model}
     \end{subfigure}
        \caption{\footnotesize Space-time plot of an incoming wave reflecting from the defect in the (a) discrete and (b) continuum models. The parameters are $d=1,K_r=2, $ and $ \gamma=0.9$. }
        \label{fig:def_ref}
\end{figure}
\begin{figure}
     \centering
     \begin{subfigure}[b]{0.45\textwidth}
         \centering
         \includegraphics[width=\textwidth]{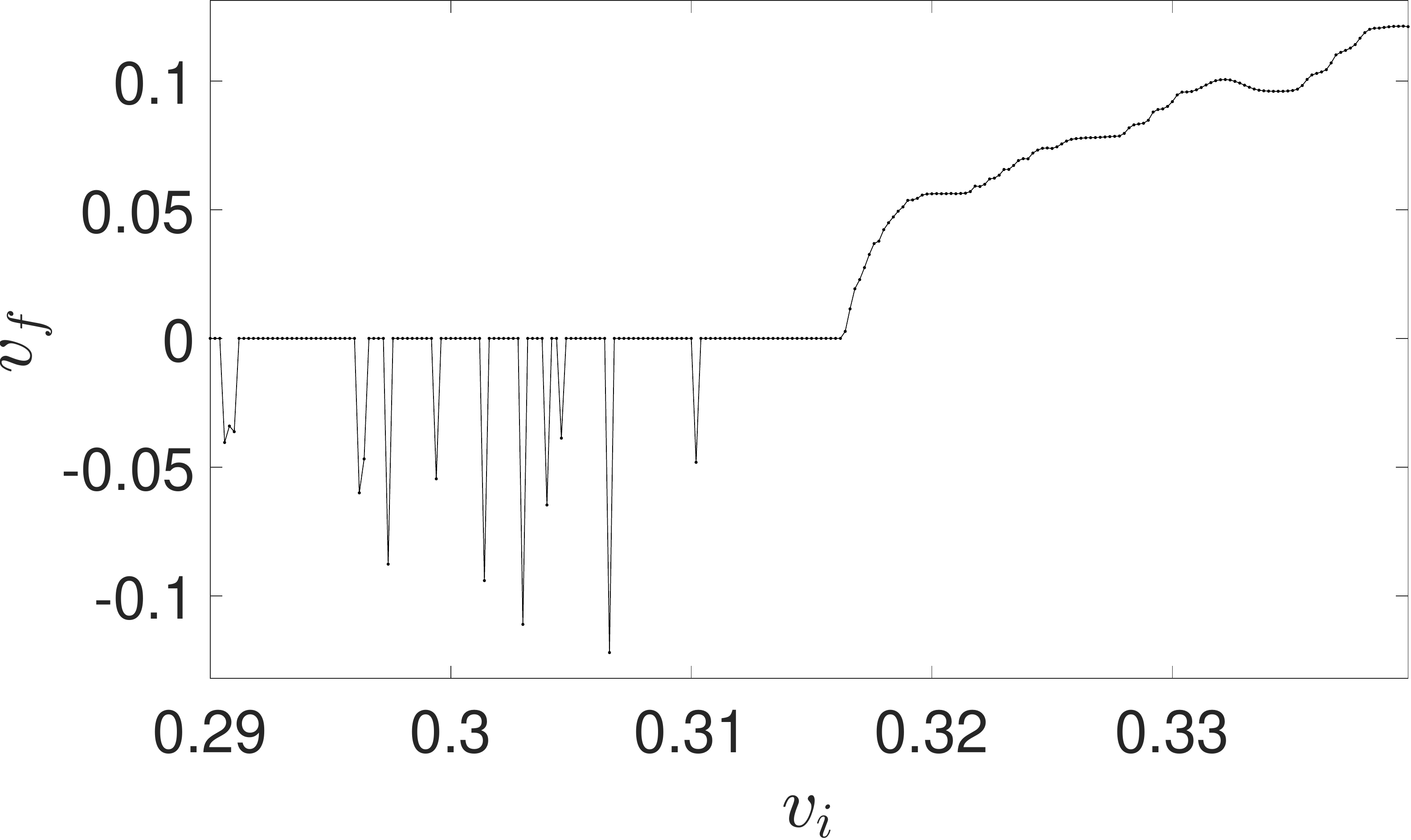}
         \caption{Discrete model}
     \end{subfigure}
     \hfill
     \begin{subfigure}[b]{0.45\textwidth}
         \centering
         \includegraphics[width=\textwidth]{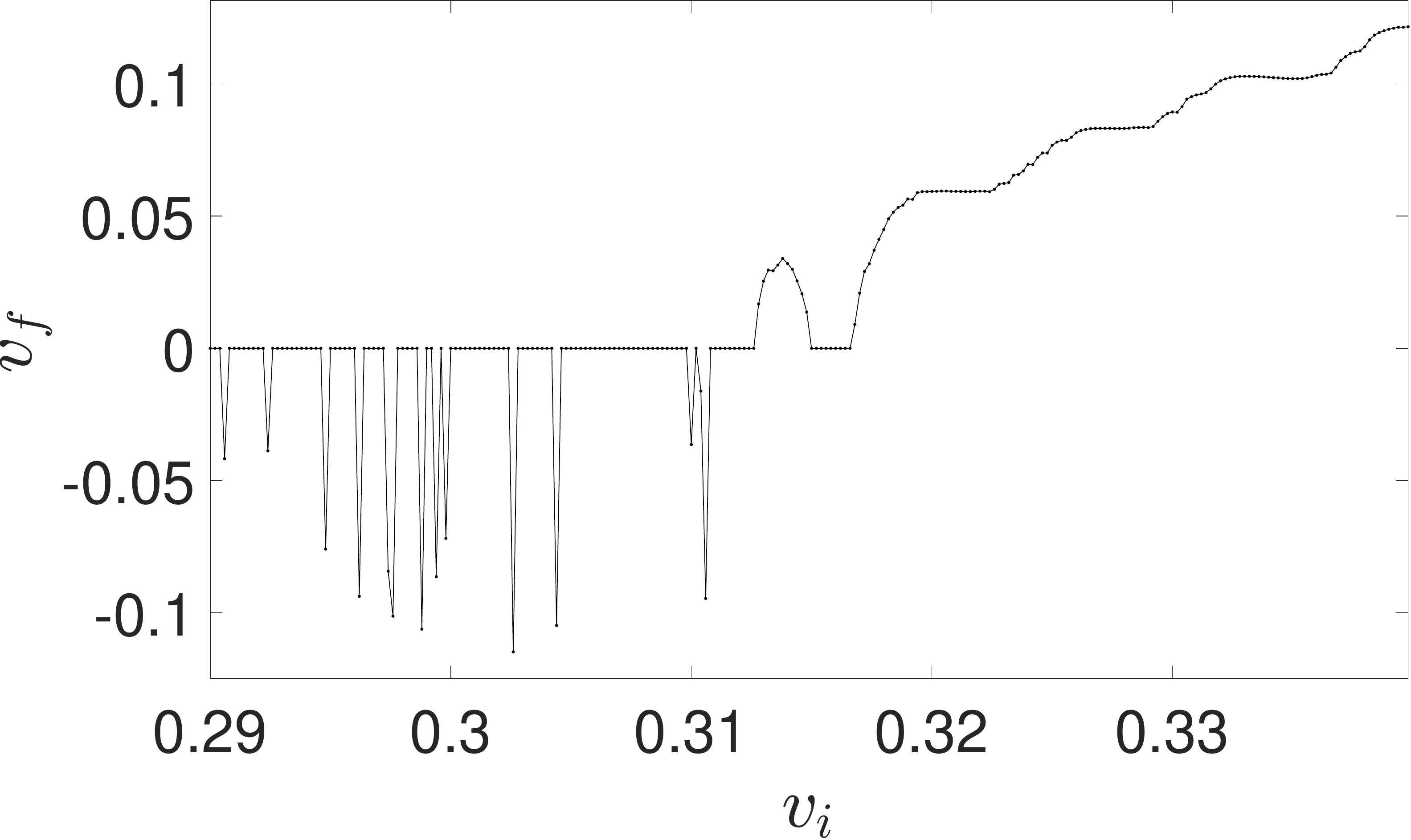}
         \caption{Continuum model}
     \end{subfigure}
        \caption{\footnotesize Input-output behavior of the system showing transmission, capture and reflection. Here $v_i$ is the velocity of an incoming solitary wave far to the left of the defect, and $v_f$ is its final velocity. The parameters are $d=1,K_r=2,\gamma=0.9$. See \cite{Note1} for a video of the three cases.}
        \label{fig:def_vivf}
\end{figure}

In Figs. \ref{fig:def_prop}, \ref{fig:def_cap} and \ref{fig:def_ref}, we show the space-time evolution of initial conditions that lead to transmission, capture, and reflection of an incoming solitary wave, respectively. We perform numerical computations using both the N-DOF system (\ref{Dimensionless_EOM}) (suitably modified to include the defect), as well as a finite difference discretization of the continuum system with defect (\ref{eq:PDE_def}). Fig. \ref{fig:def_vivf} summarizes the input-output behavior of the system with defect. If the initial speed of an incoming solitary wave, $v_i$, is equal to or below a critical velocity $v_{cr}$, it is either captured at the defect site ($v_f\approx 0$), or reflected back ($v_f<0$). For $v_i>v_{cr}$, the wave passes through the defect. In all our simulations, we use a fully formed solitary wave profile (far to the left of the defect) as an initial condition to avoid phonon excitation (`tingling') in the discrete system. Reflection and/or trapping of incoming solitary waves has been reported in earlier studies involving graded stiffness in 1D chains \cite{hwang2018solitary}, and localized defects in 2D structures supporting vector solitary waves \cite{deng2018metamaterials}.
\section{Reduced order model}
\subsection{Derivation}
To understand the numerical results discussed in the previous section, we derive a reduced order model for the system using the method of collective coordinates \cite{dauxois2006physics}, alternatively known as the reduced Lagrangian approach \cite{perez1996low}. In this approach, an ansatz is chosen for the solution, and the Euler-Lagrange equations are obtained by restricting the principle of stationary action among the class of solutions representable by that ansatz. Usually, the known exact or approximate coherent structures are included in the ansatz. Following \cite{goodman2002interaction}, we use an ansatz that assumes that the spatial profiles of the solitary wave $\hat{u}_k$ and the breather $\hat{u}_b$ are unaffected by their interaction. Specifically, we pick the ansatz
\begin{equation}\label{u=uk+ub}
u(x,X(t),a(t))=u_{k}(x,X(t))+u_{b}(x,a(t)),
\end{equation}
where $X(t)$ is location of the kink, $u_k=\hat{u}_k(x-X(t))$, $a(t)$ is the amplitude of the breather, and $u_b=a(t)\phi(x)$. 

\begin{figure}
 \centering
 \includegraphics[width=.45\textwidth]{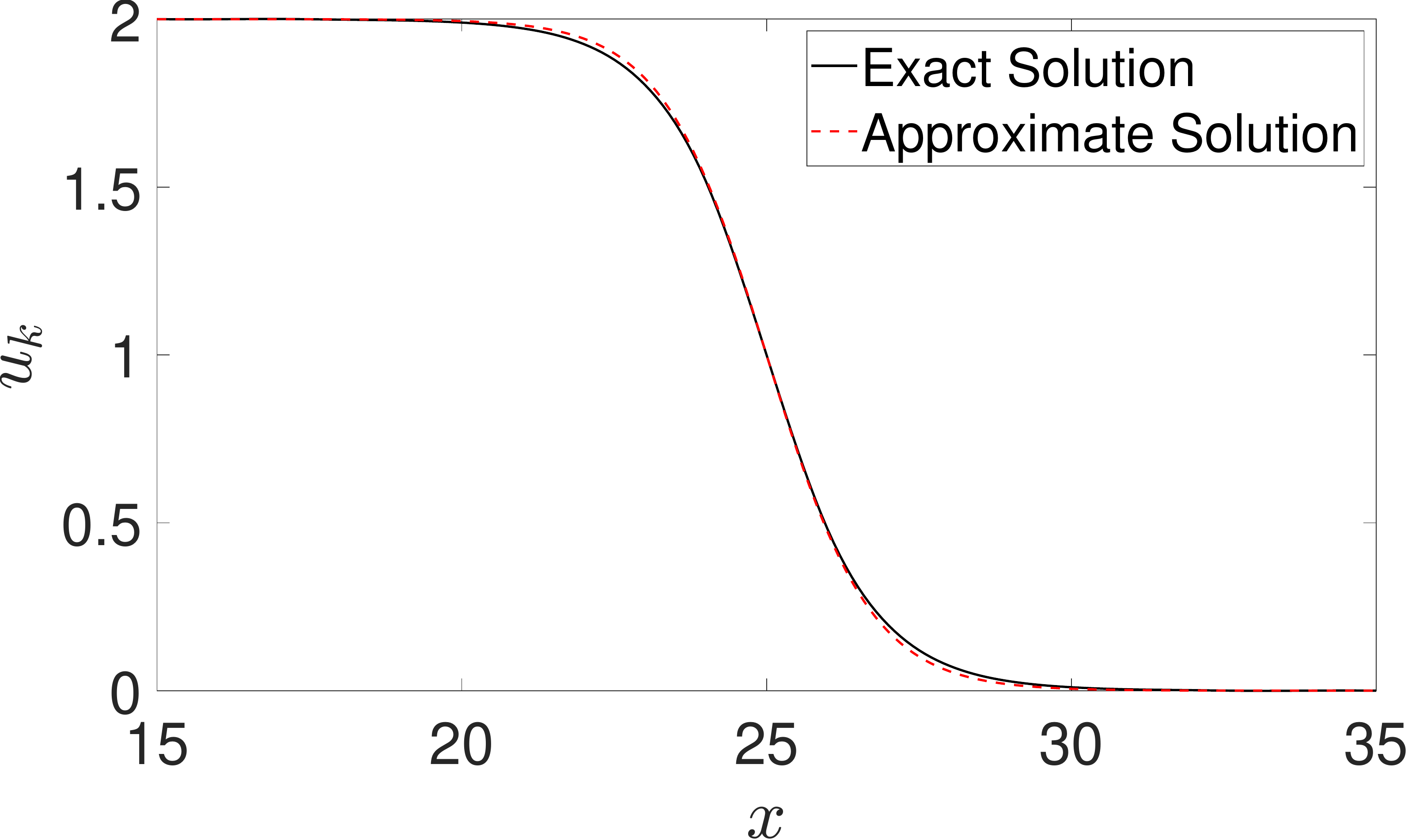}
\caption{\footnotesize Exact (solid) and approximate (dashed) shape of the solitary wave. The exact solution is obtained by numerically solving (\ref{eq:exact_kink}), and has a slope $\dfrac{du_k}{dz}\approx\dfrac{\sqrt{2}}{\sqrt{C_0^2-v^2}}(d-\sqrt{1+d^2})$ at $u_k=1$ with $z_0=25$. The approximate solution is given by (\ref{eq:approx_kink}), where $C_2$ is chosen to match the slope of the exact solution at $u_k=1$ with $X_0=25$. (for $d=1$, $z_0=25$, $v=1$, and $K_r=2$) }
 \label{fig:Kink_Exact_Approx}
 \end{figure}

The Lagrangian for (\ref{eq:PDE_def}) is
\begin{equation}\label{Lagrangian}
L=\int_{-\infty}^{\infty}\left[\frac{1}{2}u_{,t}^2-\frac{1}{2}K_{r}u_{,x}^2-\left(1-\gamma\delta(x)\right)\psi(u)\right]\,dx.
\end{equation}
Taking the derivative of (\ref{u=uk+ub}) with respect to $x$ and $t$ yields
\begin{equation}\label{ux,ut}
	u_{,x}=u_{k,x}+u_{b,x},u_{,t}=u_{k,t}+u_{b,t}
\end{equation}
where 
\begin{equation}\label{ukz}
	u_{k,t}=-\dot{X}\;u_{k,z},  u_{k,x}=\;u_{k,z}, u_{b,x}=a\phi_{,x}, \text{ and } u_{b,t}=\dot{a}\phi.
\end{equation}
We approximate higher powers of the derivatives as follows
\begin{equation}\label{ut^2}
	u_{,t}^{2}=(u_{k,t}+u_{b,t})^{2}\approx u_{k,t}^{2}+u_{b,t}^{2},
\end{equation}
and
\begin{equation}\label{ux^2}
	u_{,x}^{2}=(u_{k,x}+u_{b,x})^{2}\approx u_{k,x}^{2}+u_{b,x}^{2}.
\end{equation}
Substituting (\ref{u=uk+ub},\ref{ut^2},\ref{ux^2}) into (\ref{Lagrangian}) yields 
\begin{equation}\label{LagDetailed}
L(X,a,\dot{X},\dot{a})=\int_{-\infty}^{\infty}\left[\frac{1}{2}u_{k,t}^2+\frac{1}{2}u_{b,t}^2-\frac{1}{2}K_{r}u_{k,x}^2-\frac{1}{2}K_{r}u_{b,x}^2-\left(1-\gamma\delta(x)\right)\psi(u_{k}+u_{b})\right]\,dx.
\end{equation}
Since there is no explicit solution of (\ref{eq:exact_kink}) for $u_{k}$, we will approximate the solitary wave as 
\begin{equation}\label{eq:approx_kink}
u_{k}(x,X)=1-\tanh\left(\frac{x-X}{\sqrt{2}C_2}\right),
\end{equation}
where $C_2=\dfrac{\sqrt{C_0^2-v^2}}{2(\sqrt{1+d^2}-d)}$ has been chosen such that the slopes of the approximate and exact solutions agree at $u_k=1$, see Fig. \ref{fig:Kink_Exact_Approx}.
Since $C_0\gg v$, we use $C_2\approx \dfrac{C_0}{2(\sqrt{1+d^2}-d)}$. The fifth integral in (\ref{LagDetailed}) can be written as
\begin{multline}\label{NonlinPotInteg}
\int_{-\infty}^{\infty}(1-\gamma\delta(x))\psi(u_{k}+u_{b})\,dx
=\int_{-\infty}^{\infty}\psi(u_{k}+u_{b})\,dx-\int_{-\infty}^{\infty}\gamma\delta(x)\psi(u_{k}+u_{b})\,dx\\
=\int_{-\infty}^{\infty}\psi(u_{k}+u_{b})\,dx-\gamma\psi(u_{k}(0,X)+a).
\end{multline}

To calculate the integral in (\ref{NonlinPotInteg}), we Taylor expand the nonlinear potential energy term $\psi(u_k+u_b)$ around $a=0$, assuming that the excitation of the breather mode is weak. The rest of the details of computation of the integrals in (\ref{LagDetailed}) are relegated to the Appendix.

The effective Lagrangian is computed by substituting (A1-A7) into (\ref{LagDetailed}), 
\begin{multline}\label{Eff_Lag}
L_{eff}(X,a,\dot{X},\dot{a})=\frac{A}{2\sqrt{2}C_{0}}\dot{X}^{2}+\frac{1}{2\kappa}\dot{a}^{2}-C_{0}^2\left(\frac{A}{2\sqrt{2}C_{0}}+\frac{\kappa}{2}a^{2}\right)-\frac{C_{0}A}{2\sqrt{2}}-\frac{a^2}{(1+d^2)\kappa}+\gamma\psi\left(u_{k}(0,X)+a\right)
\\=\frac{A}{2\sqrt{2}C_{0}}\dot{X}^{2}+\frac{1}{2\kappa}\dot{a}^{2}-\left(\frac{1}{(1+d^2)\kappa}+\frac{C_{0}^2\kappa}{2}\right)a^{2}-\frac{C_{0}A}{\sqrt{2}}+\gamma\left(R(a)+F(X,a)+G(X)\right),
\end{multline} where
\begin{multline}
\label{F}
F(X,a)=2\left\{a\;\tanh\left(\frac{X}{\sqrt{2}\;C_2}\right)-\sqrt{1+d^2}\sqrt{\left(\tanh\left(\frac{X}{\sqrt{2}\;C_2}\right)+a\right)^2+d^2}\right.\\+\left.\tanh^{2}\left(\frac{X}{\sqrt{2}\;C_2}\right)-1\right\},
\end{multline}
\begin{align}\label{G}
G(X)=1-\tanh^{2}\left(\frac{X}{\sqrt{2}\;C_2}\right),\\
R(a)=a^{2}+2(1+d^{2}),
\end{align}
and
\begin{equation}\label{A}
 A=2\sqrt{1+d^{2}}-d^{2}\ln\left[\frac{\sqrt{1+d^{2}}+1}{\sqrt{1+d^{2}}-1}\right].\end{equation}
The above choice of $F(X,a)$ and $G(X)$ is crucial for the perturbation theory arguments that we employ later in the paper. 
The equations of the motion derived from the effective Lagrangian are

\begin{align}
\frac{A}{\sqrt{2}C_{0}}\ddot{X}-\gamma\left(\frac{\partial F}{\partial X}+\frac{dG}{dX}\right)=0,\label{eq:LagEOM_X}\\
\frac{1}{\kappa}\ddot{a}+\left(C_{0}^2 \kappa+\frac{2}{(1+d^2)\kappa}\right)a^{2}-\gamma\left(\frac{\partial F}{\partial a}+\frac{dR}{da}\right)=0.\label{eq:LagEOM_a}
\end{align}
The expressions for partial derivatives in the above equations are provided in the Appendix. The system in (\ref{eq:LagEOM_X}, \ref{eq:LagEOM_a}) is a conservative two degree of freedom system governing the evolution of the position of the centre of the solitary wave, $X$, and the breather amplitude, $a$.

\subsection{Numerical Simulations}
\begin{figure}
     \centering
        \begin{subfigure}[b]{0.45\textwidth}
         \centering
         \includegraphics[width=\textwidth]{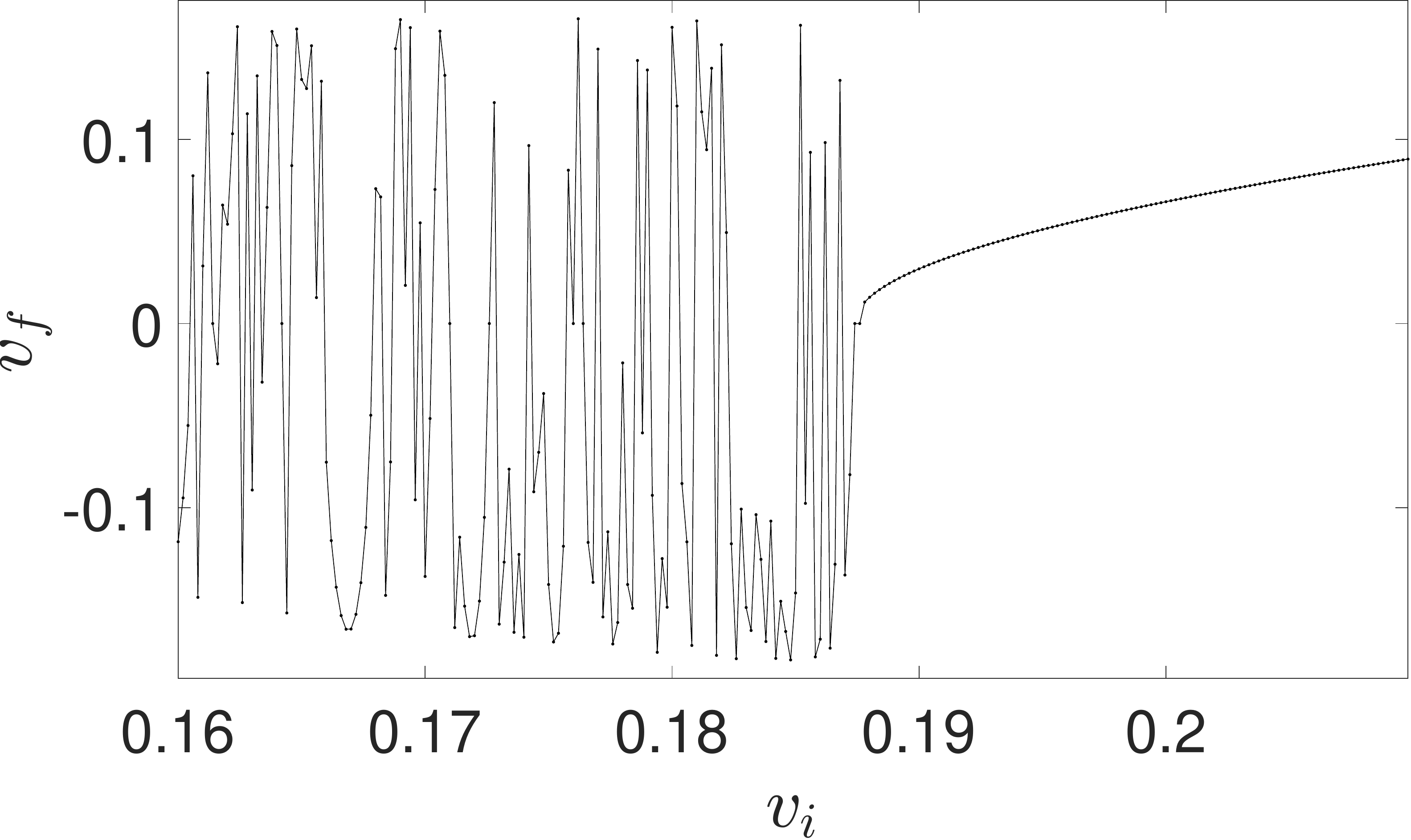}
         \caption{}
     \end{subfigure}
     \hfill
  \begin{subfigure}[b]{0.45\textwidth}
         \centering
         \includegraphics[width=\textwidth]{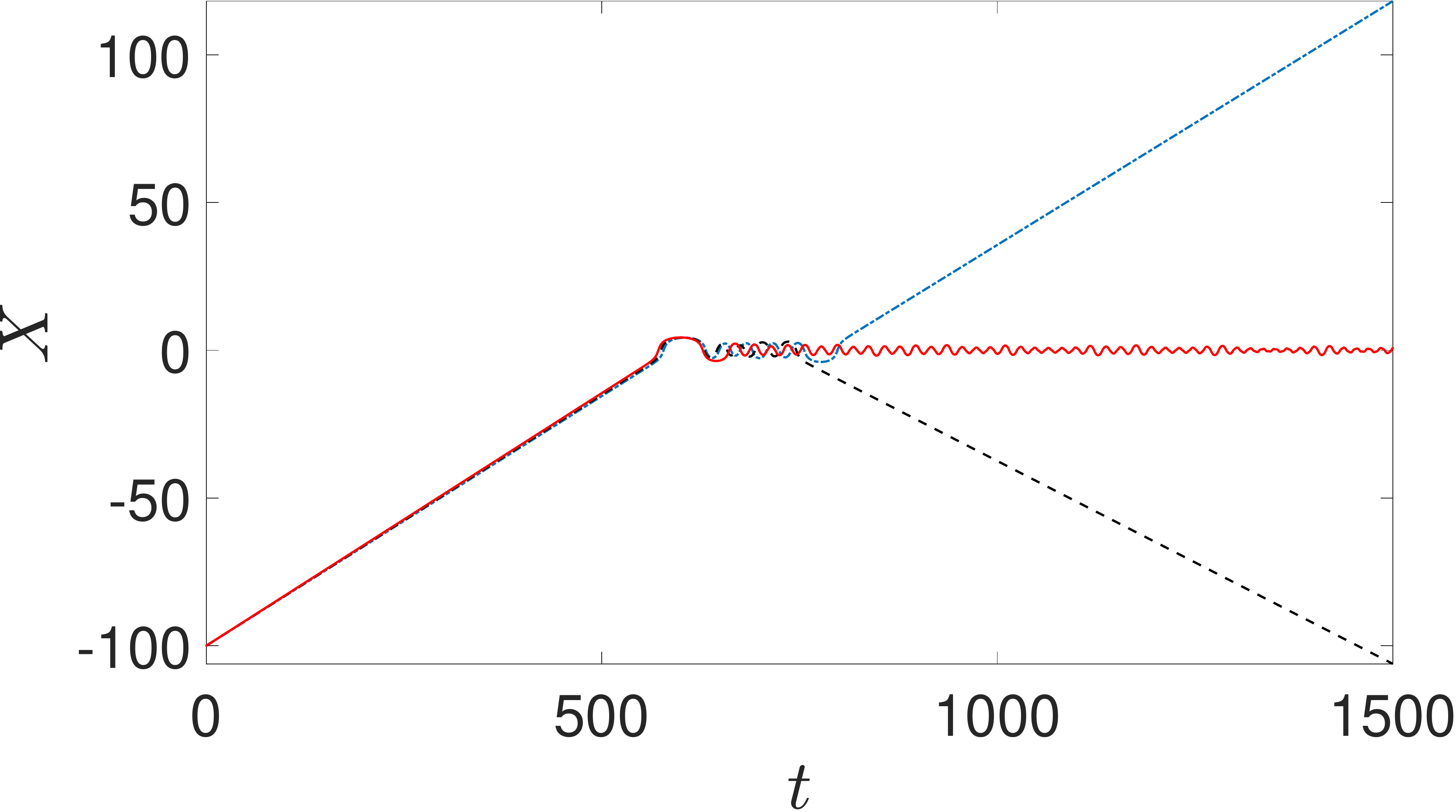}
          \caption{}
     \end{subfigure}
        \caption{\footnotesize (a) Input-output behavior of the reduced-order system with $\gamma = 0.9$. Here, $v_f$ and $v_i$ denote final and initial wave velocities. (b) Capture (bold), reflection (dash) and transmission (dot-dash) of incoming solitary waves in the reduced-order model with initial velocities (0.171,0.170,0.169), respectively.  }
        \label{fig:ro_vivf}
\end{figure}

In Fig. \ref{fig:ro_vivf}(a), we show the input-output behavior obtained by solving the 4D reduced-order dynamical system given by (\ref{eq:LagEOM_X}, \ref{eq:LagEOM_a}). To mimic the initial conditions used in simulations of the full-order models in Section II, the initial conditions $(X=-100,a=0,\dot{a}=0)$ are kept fixed, and the initial solitary wave speed $\dot{X}(0)>0$ is varied. Fig. \ref{fig:ro_vivf}(b) shows the time evolution for three different initial conditions that lead to capture, transmission and reflection, respectively. 

For $\gamma=0.9$, the critical velocity obtained using this model is  $v_{cr}\approx 0.19$, which is about $35\%$ lower than the critical velocity for full-order models. Contrary to the full-order models, capture is rarely seen in the reduced model, and there exist several intervals of initial velocity below $v_{cr}$ that lead to transmission upon interaction with the defect. In the next section, we interpret these results by analyzing phase space transport in the reduced-order dynamical system.



\section{Phase Space Analysis of the Reduced Model}
Following the approach of \cite{goodman2002interaction}, we analyze the reduced order model using a perturbative approach. To do this, we introduce a perturbation parameter $\mu$ that is a measure of the coupling between the solitary wave and the breather dynamics in the system. We use lobe dynamics and Melnikov theory to understand phase space transport in the 4D dynamical system in the limit of small (but non-zero) coupling. We show that for $\mu\ll1$, there exist heteroclinic orbits that correspond to solitary waves that transmit across the defect with vanishing initial and final speeds. In this limit, we prove the existence of chaotic dynamics, and provide a phase space interpretation of the transmitting and reflecting trajectories, as well as that of the critical velocity. Finally, we argue that the qualitative picture persists for the fully coupled case of $\mu=1$, and compute the corresponding heteroclinic orbits.
\subsection{Hamiltonian Formulation} 
From the expression of the effective Lagrangian in (\ref{Eff_Lag}), we obtain the Hamiltonian as
\begin{equation}\label{General_Hamiltonian}
H(X,a,p_X,p_a)=\dot{X}p_{X}+\dot{a}p_{a}-L_{eff}
\end{equation}
where $p_{X}$ and $p_{a}$ are the momentum variables corresponding to the collective coordinates $X$ and $a$, respectively. These momenta are computed as follows
\begin{align}
p_{X}=\frac{\partial L_{eff}}{\partial \dot{X}}=\frac{A}{\sqrt{2}C_{0}}\dot{X}\label{Px},\\
p_{a}=\frac{\partial L_{eff}}{\partial \dot{a}}=\frac{1}{\kappa}\dot{a}\label{Pa}.
\end{align}
Substituting (\ref{Eff_Lag},\ref{Px},\ref{Pa}) into (\ref{General_Hamiltonian}) yields
\begin{multline}\label{eq:H}
H(X,a,p_X,p_a)=\frac{\sqrt{2}C_{0}}{2A}P_{X}^{2}+\frac{\kappa}{2}P_{a}^{2}+\left(\frac{1}{(1+d^2)\kappa}+\frac{C_{0}^2\kappa}{2}\right)a^{2}-\gamma\left(R(a)+F(X,a)+G(X)\right)+\frac{C_{0}A}{\sqrt{2}}.
\end{multline}

\subsection{Perturbation Analysis}

We introduce a coupling parameter $0\leq\mu\leq1$ to apply perturbation theoretic arguments. The new Hamiltonian is taken to be
\begin{multline}\label{eq:Hper}
H(X,a,p_X,p_a)=\frac{\sqrt{2}C_{0}}{2 A}p_{X}^{2}+\frac{\kappa}{2}p_{a}^{2}+\left(\frac{1}{(1+d^2)\kappa}+\frac{C_{0}^2\kappa}{2}\right)a^{2}-\gamma\left(R(a)+\mu F(X,a)+G(X)\right)+\frac{C_{0}A}{\sqrt{2}}.
\end{multline}
\subsubsection{Uncoupled Case ($\mu=0$)}
For $\mu=0$, the $X$ and $a$ dynamics are uncoupled, and the Hamiltonian can be written as (ignoring constant terms)
\begin{equation}\label{eq:uncplH}
H=H^{X}+H^{a},
\end{equation}
where
\begin{align}
H^{X}=\frac{\sqrt{2}C_{0}}{2 A}p_{X}^{2}-\gamma G(X),\label{eq:uncplHx}\\
H^{a}=\frac{\kappa}{2}p_{a}^{2}+\left(\frac{1}{(1+d^2)\kappa}+\frac{C_{0}^2\kappa}{2}\right)a^{2}-\gamma R(a)\label{Ha}.\end{align}

\begin{figure}     \centering
 \begin{subfigure}[b]{0.485\textwidth}
      \centering
\includegraphics[width=\textwidth]{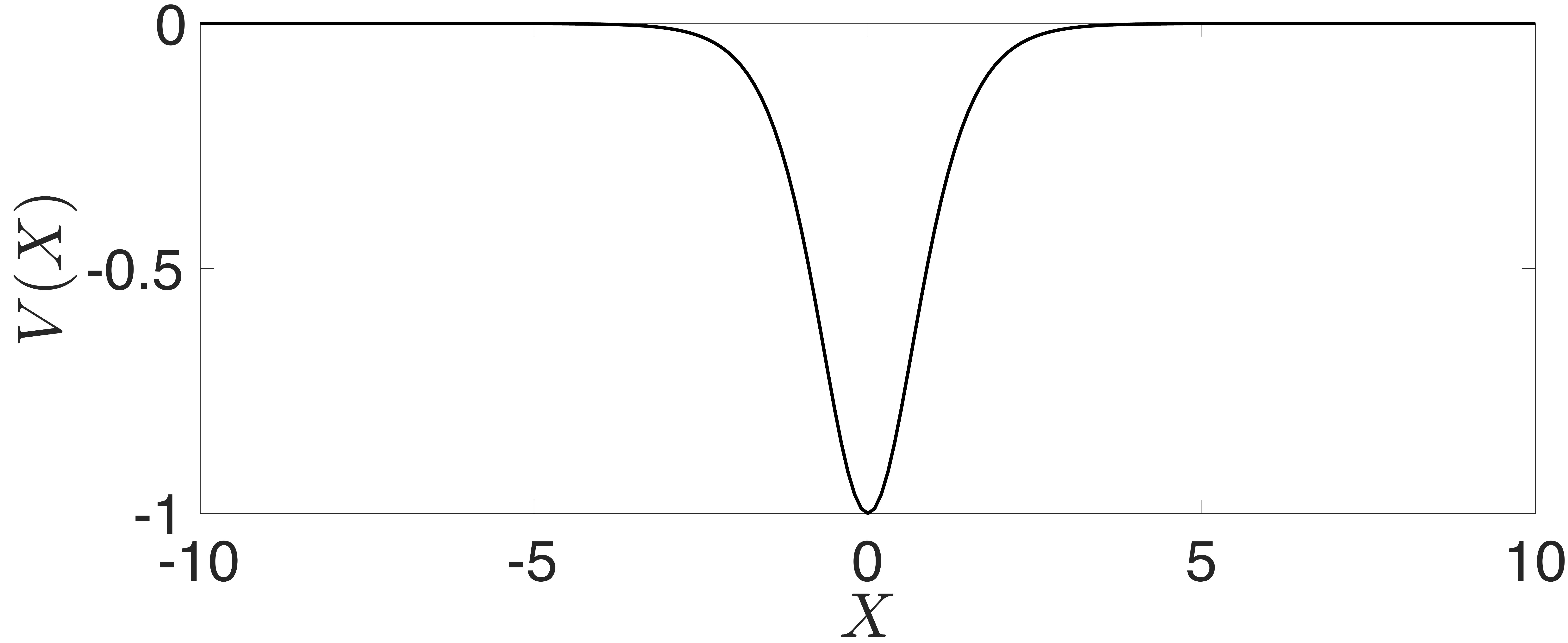}\label{}
 \end{subfigure}\hfill
 \begin{subfigure}[b]{0.485\textwidth}
      \centering

\includegraphics[width=\textwidth]{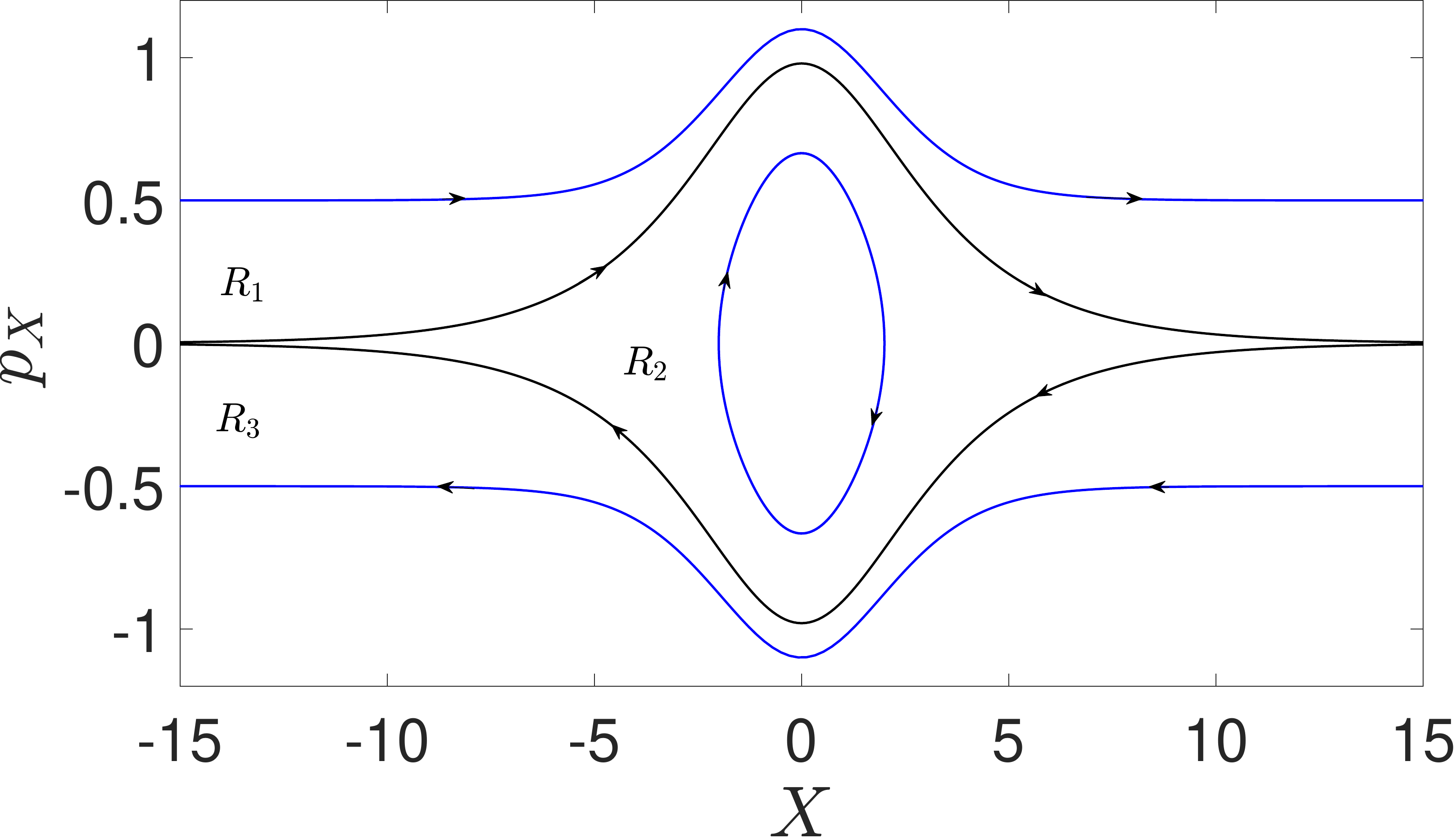}\label{}
 \end{subfigure}
 \caption{\footnotesize The $X$ dynamics in the uncoupled case ($\mu=0$). (a) Potential Energy. (b) The two heteroclinic orbits connecting the fixed points at $X=\pm\infty$, with parameters  $d=1$, $K_r=2$,and $\gamma = 0.9$. The region $R_2$ is enclosed by these two orbits. Also shown are typical trajectories in the three regions $R_1,R_2,R_3$.}
\label{fig:Uncoupled_Heteroclinc}
\end{figure}

Using Hamilton's equations, we get from (\ref{eq:uncplHx})
\begin{align}\label{eq:uncplX}
\dot{X}=\frac{\partial H^{X}}{\partial p_{X}}=\frac{\sqrt{2}C_{0}}{ A}p_{X},\\
\dot{p_{X}}=-\frac{\partial H^{X}}{\partial X}=\gamma\frac{dG}{dX}=-\frac{\sqrt{2} \gamma}{C_2}\;\sech^{2}\left(\frac{X}{\sqrt{2}\;C_2}\right)\; \tanh\left(\frac{X}{\sqrt{2}\;C_2}\right)\label{eq:uncplpX}.
\end{align}
Hence, the solitary wave dynamics are that of a particle moving under a potential $V(X)\propto -\gamma G(X)=\gamma\left(\tanh^{2}\left(\frac{X}{\sqrt{2}\;C_2}\right)-1\right)$ that has a single minima at the origin, and goes to zero as $x\rightarrow\pm\infty$. The system of equations (\ref{eq:uncplX}, \ref{eq:uncplpX}) has three fixed points $(X^*,p_X^*)$: a nonlinear center $(0,0)$, and two parabolic points $(\pm\infty,0)$.
There exist two heteroclinic orbits connecting the fixed points at $X=\pm\infty$, see Fig. \ref{fig:Uncoupled_Heteroclinc}. The phase space is divided into three disjoint regions $(R_1,R_2,R_3)$, corresponding to trajectories that are right moving, travelling on closed curves about the origin, and left moving, respectively.

 From Eqs. (\ref{eq:uncplX},\ref{eq:uncplpX}), we get
\begin{equation}\label{Xdd_uncoupled}
\ddot{X}=-\frac{2 \gamma\; C_0}{A\; C_2 } \;\sech^{2}\left(\frac{X}{\sqrt{2}\;C_2}\right)\; \tanh\left(\frac{X}{\sqrt{2}\;C_2}\right),
\end{equation} which can be integrated to obtain the equation of the heteroclinics:
\begin{equation}\label{eq:het}
X^0_{\pm}=\pm\sqrt{2}\;C_2 \sinh^{-1}\left(\frac{1}{\;C_2} \sqrt{\frac{\sqrt{2}\;\gamma\; C_0}{A}}(t-t_0)\right),
\end{equation}
where we have assumed that the two trajectories reach the origin at $t=t_0$.
From (\ref{Ha}), the governing equations for the breather are
\begin{align}\label{eq:ucpla}
\dot{a}=\frac{\partial H^{a}}{\partial p_{a}}=\kappa p_{a},\\
\dot{p_{a}}=-\frac{\partial H^{a}}{\partial a}=-(C_{0}^{2}\kappa +\frac{2}{(1+d^2)\kappa}-2\gamma)a \label{eq:ucplpa},
\end{align}
which yields $\ddot{a}+(C_{0}^{2}\kappa^2 +\dfrac{2}{(1+d^2)}-2\kappa\gamma)a=0.$
We pick parameters such that $C_{0}^{2}\kappa^2 +\dfrac{2}{(1+d^2)}-2\kappa\gamma>0$, and hence the breather mode is a linear oscillator in the uncoupled case, see the Appendix for more details.

\subsubsection{Coupled Case ($\mu>0$)}  
For $\mu>0$, the Hamiltonian is given by (\ref{eq:Hper}).
The Hamilton's equations are
\begin{align}\label{eq:cplX}
\dot{X}=\frac{\partial H}{\partial p_{X}}=\frac{\sqrt{2}C_{0}}{ A}p_{X},\\
\dot{p_{X}}=-\frac{\partial H}{\partial X}=\gamma\left(
\frac{dG}{dX}+\mu \frac{\partial F}{\partial X}\right),\label{eq:cplpX}\\
\dot{a}=\frac{\partial H}{\partial p_{a}}=\kappa p_{a}, \label{eq:cpla}\\
\dot{p_{a}}=-\frac{\partial H}{\partial a}=-(C_{0}^{2}\kappa +\frac{2}{(1+d^2)\kappa}-2\gamma)a+\gamma\mu  \frac{\partial F}{\partial a} \label{eq:cplpa}.
\end{align}

This 4D coupled system has three fixed points $(X^*,p_X^*,a^*,p_a^*)$: the origin $(0,0,0,0)$, and $(\pm \infty,0,0,0)$. For $\mu=0$, the origin is clearly a fixed point of the type centre $\times$ center. In the Appendix, we provide conditions on parameters $C_0$ and $d$ such that the origin continues to be a centre $\times$ center fixed point for all $0\leq\mu\leq 1$. This is important to ensure the validity of perturbation theory arguments that follow. 

For total energy slightly above that of the fixed points at $X=\pm\infty$, there exist periodic orbits around them. This is because in the limit $(X\rightarrow\pm\infty,p_X=0)$, the $X$ and $a$ dynamics decouple, and the pair $(a=0,p_a=0)$ is a nonlinear center of the system given by (\ref{eq:cpla}, \ref{eq:cplpa}).
\vspace{-0.3in}
\subsubsection{Poincar\'e Map and Action-Angle Coordinates}
We transform the the collective coordinate pair $(a-p_a)$ into action-angle coordinates $(I-\theta)$ via
\begin{equation}\label{I-theta coodinates}
a=S\sqrt{\omega I}\cos(\theta),\text{ and }
p_a=\sqrt{\frac{2\;\omega I}{\kappa}}\sin(\theta),
\end{equation}  
where $\omega$ is the breather frequency defined in (\ref{eq:omega}), and 
$S=\sqrt{\dfrac{2\kappa}{C_{0}^2\kappa^2+\dfrac{2}{(1+d^2)}-2\kappa\gamma}}.
$
The Hamiltonian in the transformed variables is
\begin{equation}\label{eq:cplH_aa}
H(X,p_X,I,\theta)=H^X(X,p_X)+\omega I+\mu\;H^{1}(X,I,\theta)+C_3,
\end{equation}  
where $H^X$ is given by (\ref{eq:uncplH}), $C_3=\frac{C_0 A}{\sqrt{2}}-2\gamma(1+d^2)$ and
\begin{multline}
H^1(X,I,\theta)=-\gamma F(X,I,\theta)
=-2\gamma\left(S\sqrt{\omega I}\cos(\theta)\;\tanh\left(\frac{X}{\sqrt{2}\;C_2}\right)+\tanh^{2}\left(\frac{X}{\sqrt{2}\;C_2}\right)-1\right.\\
\left.-\sqrt{1+d^2}\sqrt{\left(\tanh\left(\frac{X}{\sqrt{2}\;C_2}\right)+S\sqrt{\omega I}\cos(\theta)\right)^2+d^2}\right).
\end{multline}
Thus, the Hamilton's equations are
\begin{align}\label{eq:X_aa}
\dot{X}=\frac{\partial H}{\partial p_X}=\frac{\sqrt{2}C_0}{A}p_X,\\
\dot{p_X}=-\frac{\partial H}{\partial X}=\gamma\frac{dG}{dX}+\mu \gamma \frac{\partial F}{\partial X}, \label{eq:pX_aa}\\
\dot{\theta}=\frac{\partial H}{\partial I}=\omega-\mu\gamma \frac{\partial F}{\partial I},\label{eq:theta_aa}\\
\dot{I}=-\frac{\partial H}{\partial \theta}=\mu \gamma \frac{\partial F}{\partial \theta}.\label{eq:I_aa}
\end{align}

The uncoupled $(\mu=0)$ equations have a family of orbits
\begin{align}
&\dot{I}=0\;\;\Rightarrow \;\;I(t)=I^0,\\
&\dot{\theta}=\omega\;\;\Rightarrow \;\;\theta(t)=\omega t+\theta^0.
\end{align}

The four dimensional phase space of the coupled system is foliated by three-dimensional constant energy manifolds. Consider the manifold defined by
\begin{equation}\label{eq:fixH}
    H(X,p_X,\theta,I)=h_0,
\end{equation} where $h_0$ is a constant. On this manifold, we define a Poincar\`e Map $P_{\theta_0}$ on the two dimensional section $\Sigma^{\theta_0}=\{(X,p_X);\theta=\theta_0,H=h_0\}$. This map is globally well-defined as long as $\dfrac{\partial{H}}{\partial{I}}>0$ along the trajectories, since in that case one can invert (\ref{eq:fixH}) to obtain $I=I(X,p_X,\theta_0;h_0)$ using the implicit function theorem. From (\ref{eq:theta_aa}), we conclude that this will hold for small enough values of the coupling $\mu$.  
\vspace{-.3in}
\subsubsection{Melnikov analysis and existence of chaotic dynamics for small $\mu$}\vspace{-0.15in}
When $\mu\ll 1$, we can transform the coupled two degree of freedom system given by (\ref{eq:X_aa}-\ref{eq:I_aa}) into a single degree of freedom periodically forced system in $(X,p_X)$ \cite{guckenheimer2013nonlinear,goodman2002interaction}. In that case, we can use Melnikov's theorem to establish transversal intersection of stable and unstable manifolds of fixed points of the Poincar\`e map defined above. The Melnikov function is
\begin{align}\label{eq:Mel}
M(\theta_0,t_0)=\int_{-\infty}^{\infty} \{H^X,H^1\}(X^0,p_X^0,t+\theta_0,I^0) dt
=\int_{-\infty}^{\infty} \left(\frac{\partial H^X}{\partial X} \frac{\partial H^1}{\partial p_X}-\frac{\partial H^X}{\partial p_X} \frac{\partial H^1}{\partial X}\right) dt,
\end{align}
 where $(X^0,p_X^0)$ is the coordinate-momentum pair corresponding to the heteroclinic trajectory of the unperturbed system given by (\ref{eq:het}). This yields
\begin{multline}
M(\theta_0,t_0)=\frac{2\gamma}{C_2}\sqrt{\frac{\sqrt{2}\gamma \; C_0}{A}} \int_{-\infty}^{\infty} \sech^{3}\left(\frac{X^0}{\sqrt{2}\;C_2}\right)
\Biggl\{S \sqrt{\omega I^0} cos(t+\theta_0)+2 \tanh\left(\frac{X^0}{\sqrt{2}\;C_2}\right)\\
-\frac{\sqrt{1+d^2}\left(S \sqrt{\omega I^0} cos(t+\theta_0)+\tanh\left(\frac{X^0}{\sqrt{2}\;C_2}\right)\right)}{\sqrt{d^2+\left(S \sqrt{\omega I^0} cos(t+\theta_0)+\tanh\left(\frac{X}{\sqrt{2}C_2}\right)\right)^2}} \Biggr\} dt.\label{eq:M1}
\end{multline}

From (\ref{eq:het}), $\sech\left(\dfrac{X^0}{\sqrt{2}\;C_2}\right)=\dfrac{1}{\sqrt{1+N^2(t-t_0)^2}}$, and  $\tanh\left(\dfrac{X^0}{\sqrt{2}\;C_2}\right)=\dfrac{N(t-t_0)}{\sqrt{1+N^2(t-t_0)^2}}$,
where $N=\frac{1}{C_2}\sqrt{\frac{\sqrt{2}C_0\; \gamma}{A}}$. Inserting these expressions into (\ref{eq:M1}), we get
\begin{multline}
M(\theta_0,t_0)=\frac{2\gamma}{C_2}\sqrt{\frac{\sqrt{2}\gamma \; C_0}{A}} \int_{-\infty}^{\infty} \frac{1}{(1+N^2(t-t_0)^2)\sqrt{1+N^2(t-t_0)^2}}\times\\
\Biggl\{S \sqrt{\omega I^0} cos(t+\theta_0)+2 \frac{N(t-t_0)}{\sqrt{1+N^2(t-t_0)^2}}\\
-\frac{\sqrt{1+d^2}\left(S \sqrt{\omega I^0} cos(t+\theta_0)+ \frac{N(t-t_0)}{\sqrt{1+N^2(t-t_0)^2}}\right)}{\sqrt{d^2+\left(S \sqrt{\omega I^0} cos(t+\theta_0)+ \frac{N(t-t_0)}{\sqrt{1+N^2(t-t_0)^2}}\right)^2}} \Biggr\} \,dt.
\end{multline}
Assuming $t_0=0$, the Melnikov function can be written as 
\begin{equation}\label{eq:Melfuntheta0}
M(\theta_0)=\frac{2\gamma}{C_2}\sqrt{\frac{\sqrt{2}\gamma \; C_0}{A}} \int_{-\infty}^{\infty} Q(t) [M_1(t,\theta_0)+M_2(t)-M_3(t,\theta_0)] dt,   
\end{equation}
where $Q(t)= \dfrac{1}{(1+N^2 t^2)\sqrt{1+N^2 t^2}},
M_1(t,\theta_0)=S \sqrt{\omega I^0} cos(t+\theta_0), 
M_2(t)=2 \dfrac{N t}{\sqrt{1+N^2 t^2}},$ and $
M_3(t,\theta_0)=\dfrac{\sqrt{1+d^2}\left(S \sqrt{\omega I^0} cos(t+\theta_0)+ \dfrac{N t}{\sqrt{1+N^2 t^2}}\right)}{\sqrt{d^2+\left(S \sqrt{\omega I^0} cos(t+\theta_0)+ \dfrac{N t}{\sqrt{1+N^2 t^2}}\right)^2}}.$ 

We claim that the Melnikov function $M(\theta_0)$ vanishes at $\theta_0=\pm\frac{\pi}{2}$. To prove this, we compute
\begin{align}
M_1(t,\frac{\pi}{2})=-S \sqrt{\omega I^0} \sin(t), \text{ and }\\
M_3(t,\frac{\pi}{2})=\frac{\sqrt{1+d^2}\left(-S \sqrt{\omega I^0} \sin(t)+ \frac{N t}{\sqrt{1+N^2 t^2}}\right)}{\sqrt{d^2+\left(-S \sqrt{\omega I^0} \sin(t)+ \frac{N t}{\sqrt{1+N^2 t^2}}\right)^2}}.
\end{align}
Note that $Q(t)$ is an even function of time, while $M_1(t,\frac{\pi}{2})$, $M_2(t)$,and $M_3(t,\frac{\pi}{2})$ are odd functions of time. It follows from (\ref{eq:Melfuntheta0}) that all the three terms of the integral vanish for $\theta_0=\frac{\pi}{2}$, since each integrand is a product of an even and an odd function. Hence, we have proved that $M(\dfrac{\pi}{2})=0$. Melnikov's theorem further requires that $\dfrac{\pi}{2}$ be a simple zero of $M$, i.e., $\dfrac{dM(\theta_0)}{d\theta_0}|_{\theta_0=\frac{\pi}{2}}\neq 0$. We provide a proof of this statement in the Appendix.
The preceding analysis establishes the existence of heteroclinic tangles and chaotic dynamics in the system for small values of $\mu$. It also implies that there exist orbits heteroclinic to the periodic orbits at $X=\pm\infty$ in the corresponding 4D phase space of (\ref{eq:Hper}).

\subsubsection{Lobe dynamics and phase space transport for small $\mu$}
We use the theory of lobe dynamics \cite{wiggins2013chaotic,goodman2002interaction} to interpret the orbits of solitary waves in this chaotic system. This theory states that phase space transport can be understood in terms of forward and backward mapping of parcels (called `lobes') bounded by segments of stable ($W^s_{\hat{p}_i}$) and unstable ($W^u_{\hat{p}_i}$) manifolds of the two fixed points, $\hat{p}_1=(-\infty,0)$ and $\hat{p}_2=(\infty,0)$, of the map $P_{\theta_0}$.

The phase space is again divided into three disjoint regions $R_1,R_2,R_3$, corresponding to solitary waves that are travelling right, (temporarily) captured at the defect, and traveling left, respectively. This division of phase space is performed by selecting the appropriate primary intersection points (`pips'). A point $\hat{q}_i$ belonging to the intersection of $W^u_{\hat{p}_j}$ and $W^s_{\hat{p}_k}$ is a pip if the segment $U[\hat{p}_j,\hat{q}_i]$ on $W^u_{\hat{p}_j}$ connecting $\hat{p}_j$ to $\hat{q}_i$, and the segment $S[\hat{q}_i, \hat{p}_k]$ on $W^s_{\hat{p}_k}$ connecting $\hat{q}_i$ to $\hat{p}_k$, intersect only at $\hat{q}_i$. We denote the pip formed by intersection of $W^u_{\hat{p}_1}$ and $W^s_{\hat{p}_2}$ at $X=0$ as $\hat{q}_1$, while the pip at the intersection of $W^u_{\hat{p}_2}$ and $W^s_{\hat{p}_1}$ at $X=0$ is denoted as $\hat{q}_2$. 

\begin{figure}
\centering
\includegraphics[width=0.9\textwidth]{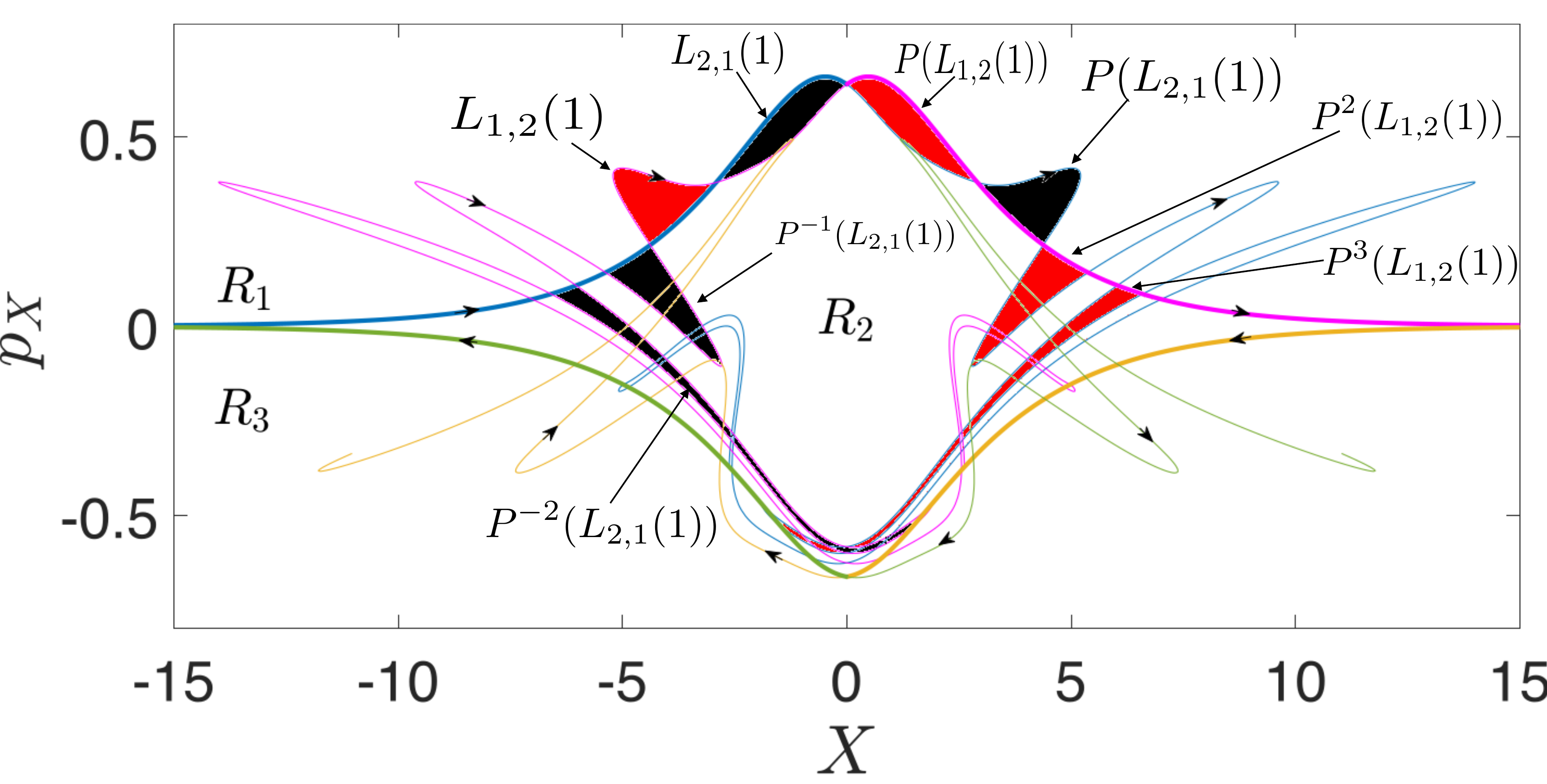}
\caption{\footnotesize The stable and unstable manifolds of the fixed points at $X=\pm\infty$ for the Poincar\`e map $P$ at the section $\theta_0=\pi/2$. The region $R_2$ is bounded by bold segments. The lobe $L_{1,2}(1)$ is the set that is mapped from  $R_1$ into $R_2$ in one iteration of $P$. This lobe and its three forward iterates are shown in red. The lobe $L_{2,1}(1)$ is the set that is mapped from $R_2$ into $R_1$ in one iteration. This lobe and its one forward as well as two backward iterates are shown in black. Note that the lobe $P^3(L_{1,2}(1))$ intersects the lobe $P^{-2}(L_{2,1}(1))$, implying that trajectories can travel from $R_1$ to $R_2$ and then back to $R_1$. The parameters are $h_0 = 0.5$, $\mu = 0.5$, and $\gamma = 0.9$.}
\label{fig:lobes_illus}
\end{figure}

Once these primary intersections points are picked, the boundaries between regions $R_i$ can be demarcated using the associated invariant manifolds, as shown in Fig. \ref{fig:lobes_illus}.  The region $R_2$ is enclosed by $U[\hat{p}_1,\hat{q}_1]$, $S[\hat{q}_1,\hat{p}_2]$, $U[\hat{p}_2,\hat{q}_2]$ and $S[\hat{q}_2,\hat{p}_1]$. The regions $R_1$ and $R_3$ are defined as $R_1=[(x,y)|(x,y>0)\notin R_2]$, and  $R_3=[(x,y)|(x,y<0)\notin R_2]$, respectively.

A lobe is an area enclosed by segments $U[\hat{q}_i,\hat{q}_j]$ and $S[\hat{q}_j,\hat{q}_i]$ for any neighboring pair of pips $\hat{q}_i$ and $\hat{q}_j$. Lobes are mapped onto each other by forward and backward iterations of the map $P_{\theta_0}$. The lobe $L_{i,j}(k)$ is the set of all points that are mapped from $R_i$ to $R_j$ after $k$ iterations of $P_{\theta_0}$. Furthermore, any point in $R_i$ that eventually enters $R_j$ must pass through $L_{i,j}(1)$. Fig. \ref{fig:lobes_illus} also shows a few forward and backward iterates of $L_{1,2}(1)$ and $L_{2,1}(1)$.



\begin{figure}
     \centering
     \begin{subfigure}[b]{0.9\textwidth}
         \centering
         \includegraphics[width=\textwidth]{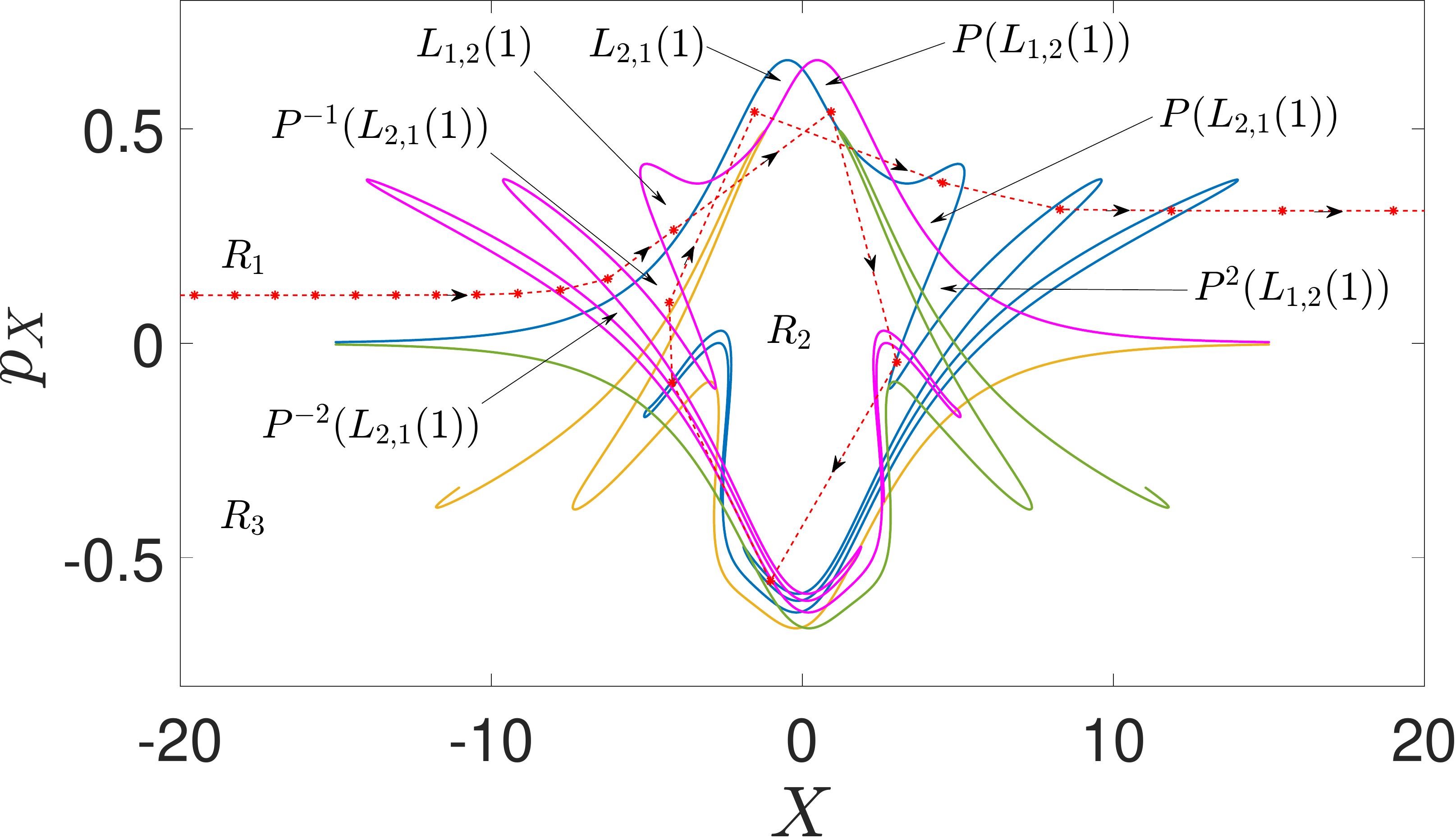}
         \caption{\footnotesize {Transmitted Wave}}
     \end{subfigure}
     \hfill
     \begin{subfigure}[b]{0.9\textwidth}
         \centering
         \includegraphics[width=\textwidth]{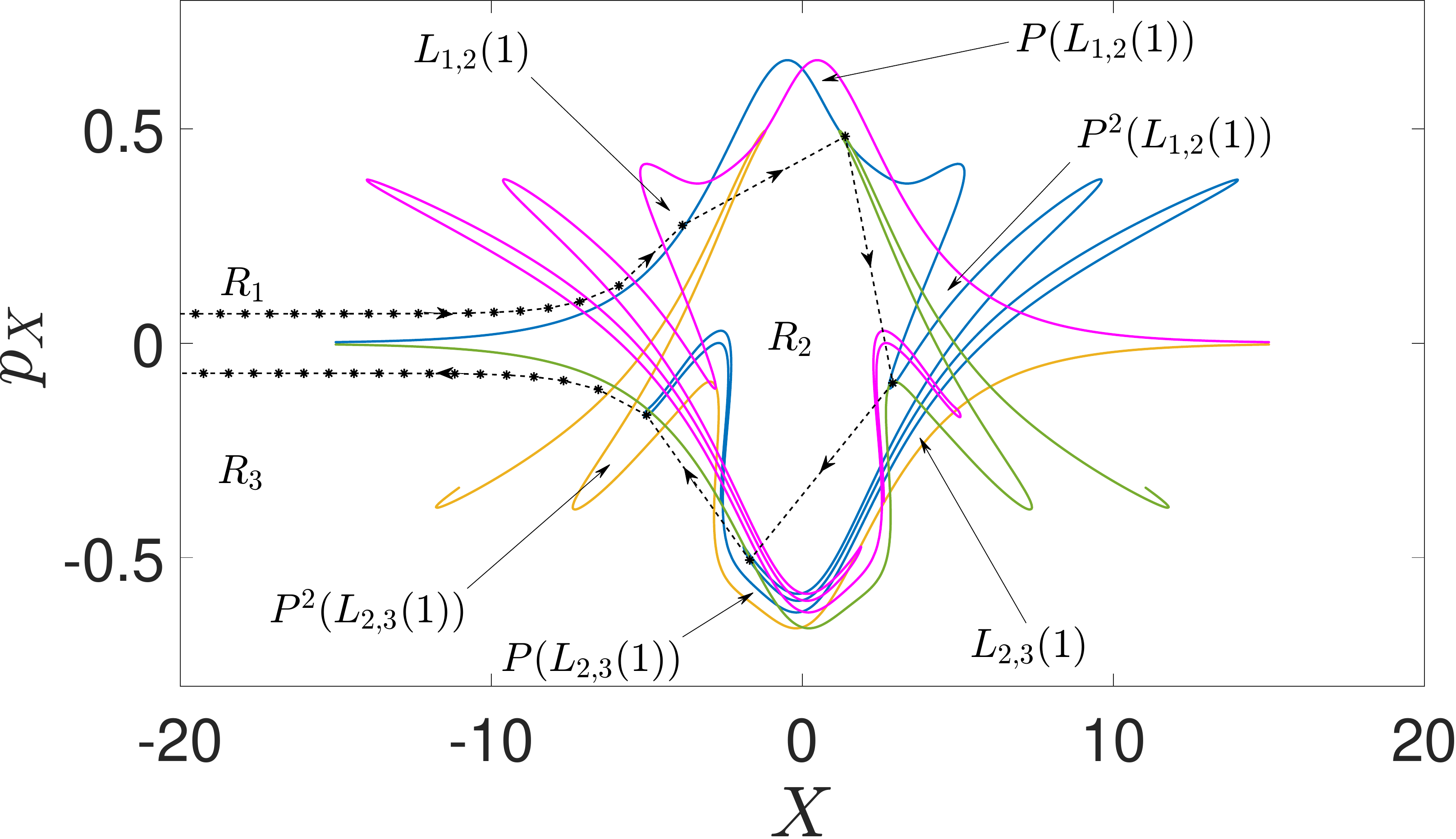}
         \caption{\footnotesize{Reflected Wave}}
     \end{subfigure}
        \caption{\footnotesize (a) Phase space evolution of a solitary wave transmitting across the defect. The trajectory begins in $R_1$, and enters $R_2$ via the lobe $L_{1,2}(1)$ bounded by $W^s_{+\infty}$ (purple) and $W^u_{-\infty}$ (blue). It is mapped onto $P^{-2}(L_{2,1}(1))$ after three iterations inside $R_2$, leading to its re-entry into $R_1$ after the sixth iteration. (b) Phase space evolution of a solitary wave reflecting back from the defect. The trajectory begins in $R_1$, and enters $R_2$ via $L_{1,2}(1)$.  After the first iteration inside $R_2$, it is mapped onto $L_{2,3}(1)$, a lobe bounded by $W^u_{+\infty}$ (orange) and $W^s_{-\infty}$ (green), leading to its entry into $R_3$ after the second iteration. The parameters are $h_0 = 0.5$, $\mu = 0.5$, and $\gamma = 0.9$.}
        \label{fig:lobes_pass_ref}
\end{figure}

For a right-moving solitary wave to transmit across the defect, it must either stay in $R_1$ for all times, or transit from $R_1$ to $R_2$, and then back to $R_1$. Fig. \ref{fig:lobes_pass_ref}(a) shows a trajectory doing the latter. Once this trajectory enters $R_2$, it is mapped onto a pre-image of $L_{2,1}(1)$, and eventually gets ejected back to $R_1$. Analogously, for a right moving solitary wave to get reflected back from the defect, it must transit from $R_1$ to $R_2$ and then from $R_2$ to $R_3$. Fig. \ref{fig:lobes_pass_ref}(b) shows such a trajectory. Once this trajectory $R_2$, it is mapped onto a pre-image of $L_{2,3}(1)$, and eventually gets ejected into $R_3$. 

Since the Poincar\`e map $P_{\theta_0}$ is area preserving, by arguments similar to those in \cite{goodman2002interaction}, the set of points that are captured by the defect for all times has measure zero. As a result, capture is observed only for isolated values of incoming velocities in the reduced order model (see Fig. \ref{fig:ro_vivf}). This is in contrast to the full-order model results in Fig. \ref{fig:def_vivf} that show existence of finite intervals of initial velocities that lead to permanent capture of incoming waves. This however does not rule out the existence of trajectories that are captured for arbitrarily long times by the defect in the reduced order model. In fact, the existence of horsehoes \cite{wiggins2003introduction} in the system leads us to conjecture that for each positive integer $n$, there exists an initial condition such that a trajectory coming into $R_2$ from $R_1$ performs $n$ clockwise `revolutions' around the origin, before exiting to either $R_1$ or $R_3$.

\subsubsection{Lobe dynamics interpretation of critical velocity}
For a given total energy level $h$, the maximum height of the sequence of lobes $P^{-n}(L_{1,2}(1))$ for ($n=1,2,3,\dots $) reaches an asymptote $p_{-\infty}$ as $n\rightarrow\infty$. If the initial momentum of an incoming solitary wave (at $X\rightarrow-\infty$) is higher than $p_{-\infty}$, then the trajectory will travel above the lobes, staying in region $R_1$ and transmitting across the defect. On the other hand, if the initial momentum of an incoming solitary wave is lower than or equal to $p_{-\infty}$, its fate will be decided by the lobe dynamics discussed previously, and all three outcomes of transmission, capture and reflection are possible.
\begin{figure}
\centering
\includegraphics[width=0.9\textwidth]{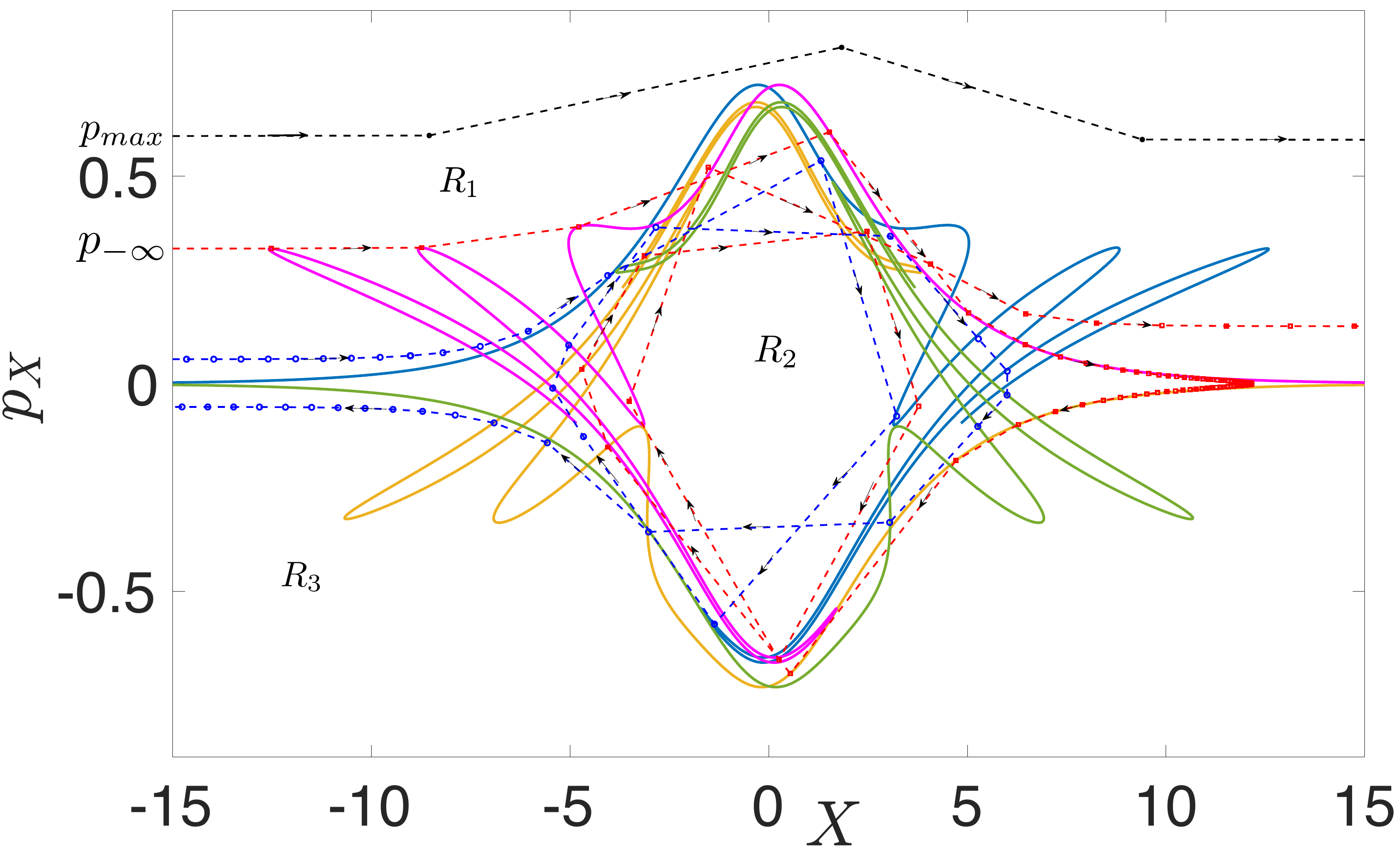}
\caption{\footnotesize{Poincar\`e section for $h=-0.4>h_{cr}=-0.72$ with $\mu = 0.5 $, and $ \gamma = 0.9$. The trajectory in black has incoming velocity $v_{i}\approx v_{max}>v_{-\infty}$, and transmits across the defect while travelling above the lobes. The trajectory in red has $v_{i}\approx v_{-\infty}$, while the one in blue has $v_{i}< v_{-\infty}$. Here $v_{-\infty}$ is the velocity corresponding to the momentum $p_{-\infty}$. The evolution of both red and blue trajectories is governed by the lobe dynamics. }}
\label{fig:lobehigh}
\end{figure}
For fixed $h$, an incoming solitary wave with maximum allowable velocity $v_{max}$ corresponds to an initial condition with no energy in the breather. Hence, $v_{max}$ can be obtained by putting $(X\rightarrow -\infty, a=0,p_a=0)$ in the Hamiltonian ($\ref{eq:cplH_aa}$), and inverting the equation $H=h$. Similarly, an incoming solitary wave with  minimum allowable velocity $v_{min}=0$ corresponds to an initial condition with all the energy in the breather. 

Recall that we defined the critical velocity $v_{cr}$ to be the velocity above which an incoming solitary wave (with zero initial energy in breather) will always transmit across the defect. Let the corresponding energy level be denoted by $h_{cr}$. Our lobe dynamics computations reveal that for any fixed energy level $h> h_{cr}$, the maximum allowable incoming momentum ($p_{max}=\dfrac{A}{\sqrt{2}C_0}v_{max}$) is higher than the corresponding $p_{-\infty}$, as shown in Fig. \ref{fig:lobehigh} for $h=-0.4$, with $h_{cr}=-0.72$. Hence for each $h> h_{cr}$, incoming waves with $v_{i}\in (\dfrac{\sqrt{2}C_0}{A}p_{-\infty},v_{max}]$ will always transmit, while those with $v_{i}\in (0, \dfrac{\sqrt{2}C_0}{A}p_{-\infty}]$ will be governed by the lobe dynamics. Fig. \ref{fig:lobehigh} shows three trajectories (all at $h=-0.4$) with incoming velocity greater than, equal to and less than $v_{-\infty}$, respectively.
\begin{figure}
\centering
\includegraphics[width=0.9\textwidth]{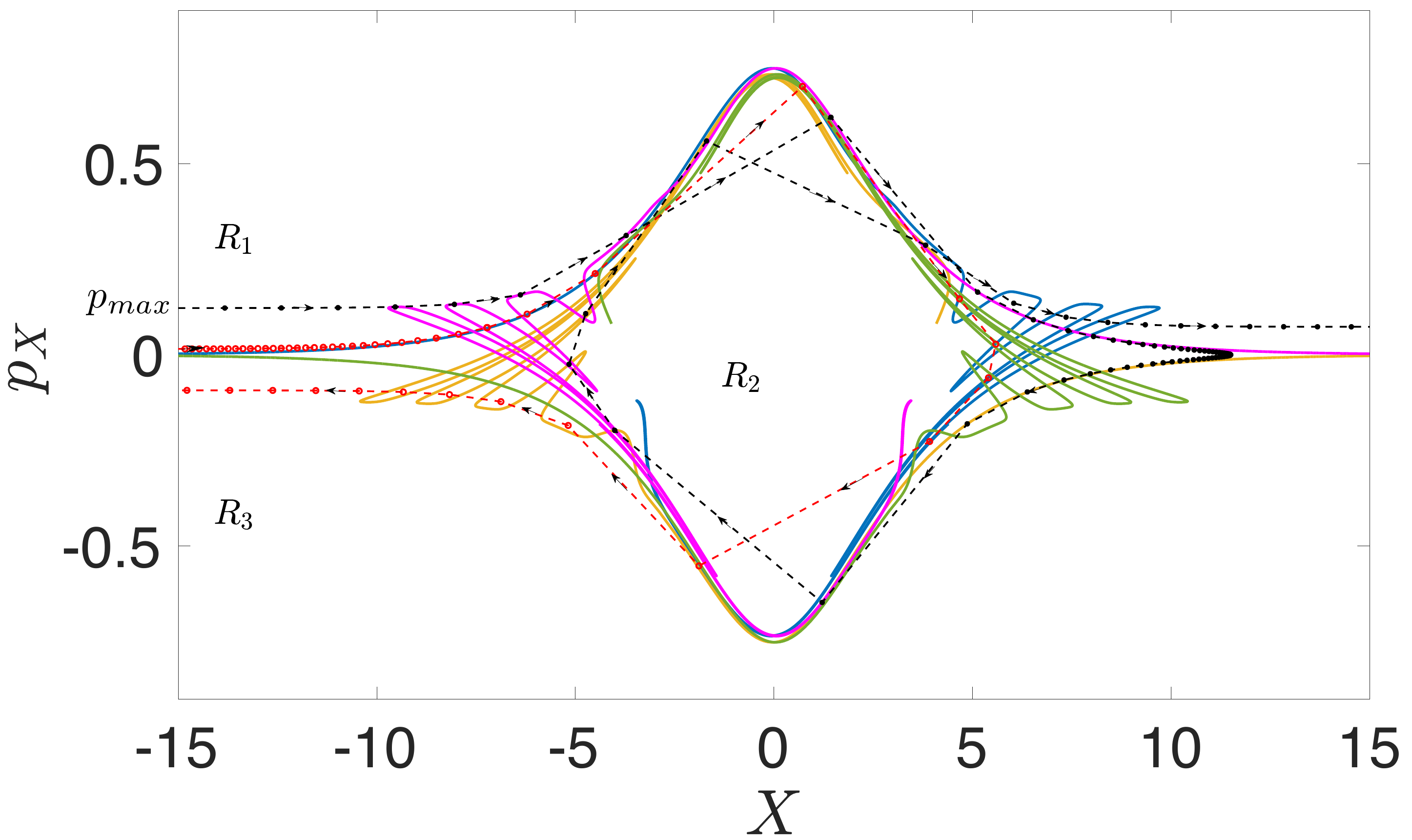}
\caption{\footnotesize{Poincar\`e section for $h\approx h_{cr}=-0.72$ with $\mu = 0.5 $, and $ \gamma = 0.9$. In this case, $p_{-\infty}\approx p_{max}$, and $v_{-\infty}\approx v_{max}$. The trajectory in black has incoming velocity $v_{i}\approx v_{max}$, while the trajectory in red has $v_{i}< v_{max}$. The evolution of both black and red trajectories is governed by the lobe dynamics. }}
\label{fig:lobelow}
\end{figure}
On the other hand, we find that for each $h\leq h_{cr}$, $p_{max}=p_{-\infty}$, and hence, the fate of all possible incoming trajectories with $v_{i}\in (0, v_{max}]$ will be governed by lobe dynamics. Fig. \ref{fig:lobelow} shows such a case with $h\approx h_{cr}=-0.72$.

To summarize, let $\Omega_1$ be set of initial conditions (at $X\rightarrow-\infty$) that lie above the $p_X=p_{-\infty}$ line, where $p_{-\infty}$ depends on the energy level of the given initial condition. We find that all trajectories originating in $\Omega_1$ travel above the lobes and always transmit across the defect. Let $\Omega_2$ be the set of initial conditions with all energy in the incoming solitary wave. Then the intersection $\Omega=\Omega_1\cap\Omega_2$ turns out to be precisely the set of initial conditions with $v_i>v_{cr}$.

\subsection{Heteroclinic orbits for $\mu=1$}
Recall that we recover the original reduced order Hamiltonian system (\ref{eq:H}) by putting $\mu=1$ in (\ref{eq:Hper}). To compute orbits heteroclinic to the periodic orbits at $X=\pm\infty$ for this case, we use the fact that the stable manifolds in Fig. \ref{fig:lobes_illus} can be obtained by reflecting the unstable manifolds across the $p_X$ axis. This is a consequence of the invariance of the system under the transformation:
\begin{equation}\label{symmetry}
(-X,p_X,a,p_a,t) \rightarrow (X,p_X,-a,p_a,-t),
\end{equation}
where $a=0$ for the Poincar\`e section of Fig. \ref{fig:lobes_illus}. Hence, if the unstable manifold of the periodic orbit at $X=-\infty$ intersects the $X=0$ surface at $a=0$, then the intersection also belongs to the stable manifold of the periodic orbit at $X=+\infty$.

Let $\phi_t(X,p_X,a,p_a)$ denote the time-$t$ flow map for (\ref{eq:Hper}). Using Matlab's BVP4C \cite{shampine2000solving}, we solve the following multi-point boundary value problem (BVP) for a trajectory beginning at $(-X^*,p_{X-},a_-,p_{a-})$ at $t=0$, reaching $(X_0,p_{X0},a_0,p_{a0})$ at $t=T/2$, and terminating at $(X^*,p_{X+}=p_{X-},a_+=-a_-,p_{a+}=p_{a-})$ at $t=T$:
\begin{align}
    \phi_{T/2}(-X^*,p_{X-},a_-,p_{a-})=(X_0,p_{X0},a_0,p_{a0}),\\
    \phi_{-T/2}(X^*,p_{X-},-a_-,p_{a-})=(X_0,p_{X0},a_0,p_{a0}).
\end{align}

To make the problem well-posed, we fix $X^*\approx 10$. The BVP consists of eight equations corresponding to the eight unknowns $(p_{X-},a_-,p_{a-},X_0,a_0,p_{X0},p_{a0},T)$. The initial guesses are obtained from heteroclinic trajectories obtained for $\mu<1$ using Poincar\`e sections, as discussed in the previous section. Once one solution to the BVP is found, we find other distinct solutions by solving the BVP with different initial guesses for $a_-$ and $p_{a-}$, while keeping the energy equal to the first solution.

\begin{figure}
     \centering
     \begin{subfigure}[b]{0.485\textwidth}
         \centering
         \includegraphics[width=\textwidth]{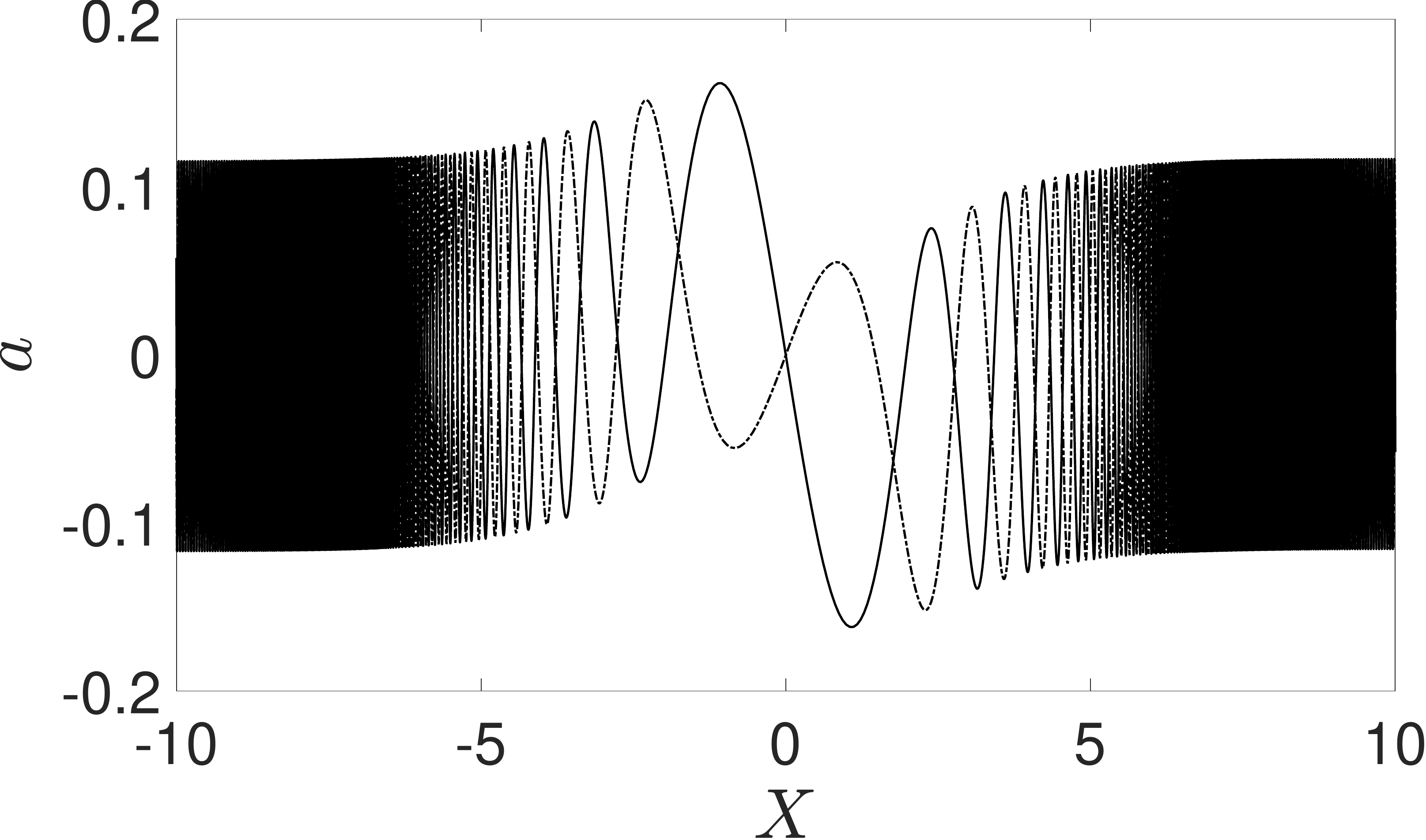}
         \caption{}
     \end{subfigure}
     \hfill
     \begin{subfigure}[b]{0.485\textwidth}
         \centering
         \includegraphics[width=\textwidth]{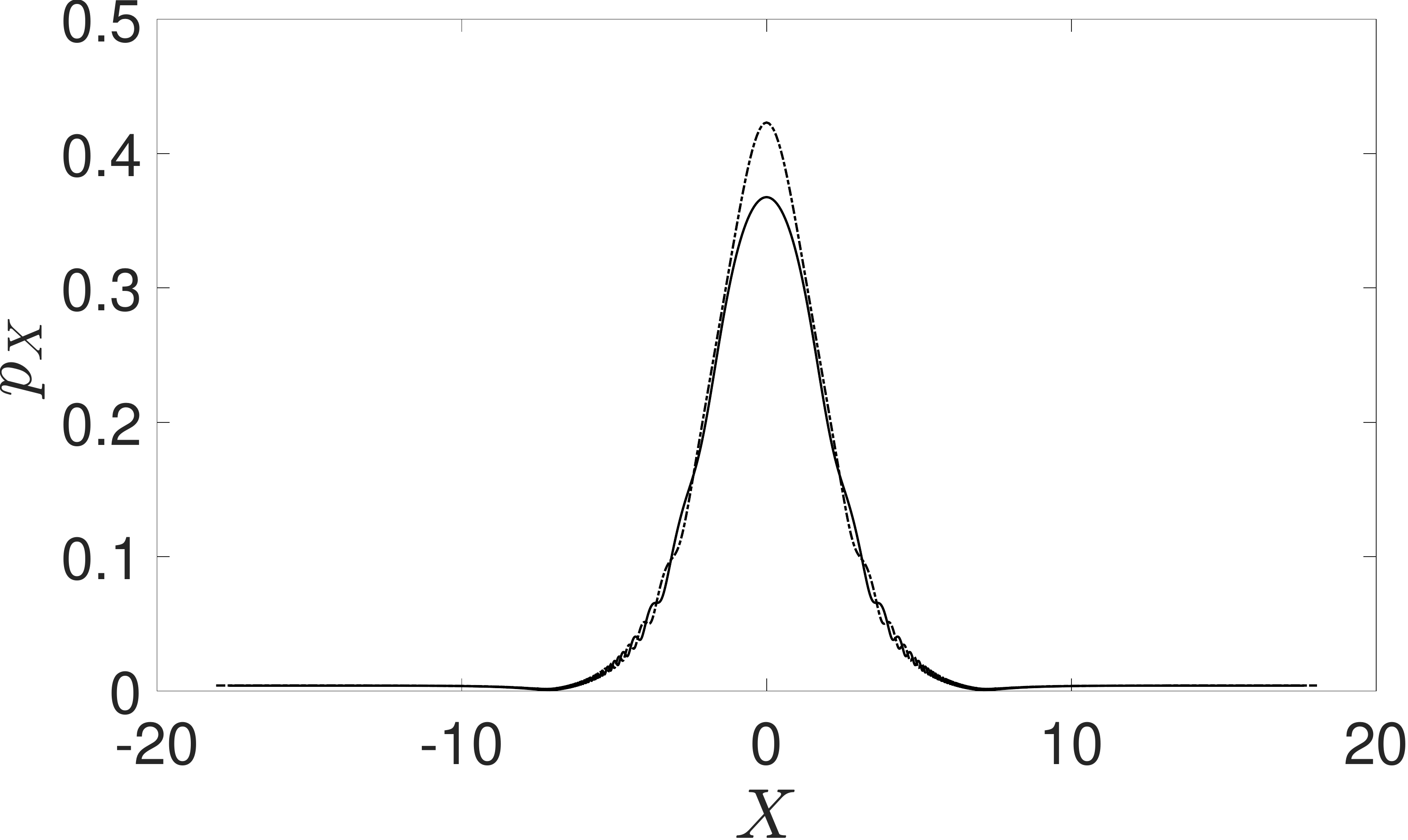}
         \caption{}
     \end{subfigure}
        \caption{\footnotesize Two distinct orbits heteroclinic to the two periodic orbits at $X=\pm\infty$, with $\mu = 1$, and $\gamma = 0.9$, projected on (a) $X-a$, and (b) $X-p_X$ planes.}
        \label{fig:mu1_het12}
\end{figure}

Fig. \ref{fig:mu1_het12} shows the projections of two such heteroclinic orbits on the $(X-a)$ and $(X-P_X)$ planes. Each right-moving heteroclinic orbit corresponds to a solitary wave that arrives from the periodic orbit at $(X\rightarrow -\infty)$ with $p_X\approx 0^+$ and all the energy initially in the breather. As this wave approaches the defect, it absorbs energy from the breather and accelerates. Once past the defect, the same amount of energy is gradually transferred back to the breather, and the wave approaches $X\rightarrow\infty$ with vanishing speed. The situation is analogous for a left-moving heteroclinic orbit.

\section{Discussion and concluding remarks}
In this paper, we have derived a reduced-order dynamical model of an infinite chain of bistable mechanical elements with a localized defect, and analyzed it using methods of dynamical systems theory. Focusing on the interactions between solitary waves and the breather mode that arises due to the defect, this two degree of freedom Hamiltonian model captures some qualitative aspects of the system dynamics.
The study of phase space transport in the reduced-order model via lobe dynamics elucidates the mechanisms via which an incoming solitary wave may get transmitted, captured, or reflected upon reaching the defect. Both the full-order and reduced-order models predict that there is a critical initial velocity (with no energy initially in the breather) above which an incoming solitary wave will always pass through the defect. However, there are a number of disagreements between the models, including the value of the critical velocity itself. While the reduced-order model predicts strong sensitivity to initial conditions below the critical velocity, the full-order models predict that an incoming wave will always be captured in that regime, except for isolated initial conditions that lead to reflection. 

These discrepancies can primarily be attributed to three factors. First, it is known that solitary wave-defect interactions lead to `leaking' of energy into the infinite dimensional subspace of linear waves (`phonons'), even in the continuum setting. Our reduced-order model can potentially be made more accurate by including this radiation damping \cite{goodman2002interaction}. It is a challenging task to derive an accurate analytical description of the damping terms to be appended to the Hamiltonian equations, since it involves a careful study of resonances between modes corresponding to the discrete and continuous spectra \cite{soffer1999resonances}. Recent advances in data-driven sparse learning of governing equations \cite{brunton2016discovering} may provide an alternative way to obtain the dominant damping terms. Once a dynamical model is available, the current framework can potentially be extended to understand the damped dynamics, since lobe dynamics and other related phase space transport methods have been used to study non-conservative systems \cite{wiggins2013chaotic,zhong2020geometry}. 

Second, prior numerical studies of solitary wave-defect interaction in the closely related $\phi^4$ model have pointed out the importance of an `internal mode' in the dynamics \cite{fei1992resonantphi4,kivshar1998internal}. This internal mode is a spatially localized eigenfunction of the system linearized about the solitary wave solution. Our computations confirm that a similar internal mode exists in the system considered in this study. It corresponds to shape change of the solitary wave profile, and it can potentially exchange energy with the solitary wave. Hence, it is plausible that including this mode into the reduced-order model could lead to a better qualitative and quantitative agreement with the results of the full-order model. 

Finally, our current reduced-order model does not take into account the effect of phonon excitation that occurs purely due to discreteness of the system. In discrete bistable systems such as those considered in this study, transition waves are accompanied by oscillatory tails \cite{hwang2018input} that vanish when discreteness parameter goes to 0. These tails consist of phonons primarily at a single frequency, and mechanisms have been proposed recently to harvest this energy by inclusion of defects  \cite{hwang2018input, hwang2019energy}. Incorporating this energy exchange into the reduced-order models is another topic for future study.

Our work is a step towards a rational approach to defect engineering in mechanical metamaterials, based on a fully nonlinear dynamical systems approach. This approach can potentially be extended to spatially extended defects, and three-DOF reduced-order models, as demonstrated in photonic metamaterials \cite{goodman2002stopping,goodman2004strong}. Another promising extension is active modulation of the defect strength (i.e., onsite spring stiffness), or the breather oscillations, in an open-loop or feedback fashion for designing active mechanical metamaterials \cite{udani2017sustaining,pishvar2020foundations,sirota2020non,pearce2020programming,rosa2020dynamics,kruss2022nondispersive}.

\appendix
\section{Details of derivation of the reduced order model}\label{app:integrals}
 The first and third integrals are
\begin{align}\label{Integ_ukt}
\int_{-\infty}^{\infty}\frac{1}{2}u_{k,t}^2\,dx=-\int_{0}^{2}\frac{1}{2}\dot{X}^{2} u_{k,z}^2\,du
=\frac{\dot{X}^{2}}{2\sqrt{2K_{r}}}(2\sqrt{1+d^{2}}-d^{2} \; ln[\frac{\sqrt{1+d^{2}}+1}{\sqrt{1+d^{2}}-1}]), \\
\int_{-\infty}^{\infty}\frac{1}{2}u_{k,x}^2\,dx=-\int_{0}^{2}\frac{1}{2} u_{k,z}^2\,du
=\frac{1}{2\sqrt{2K_{r}}}(2\sqrt{1+d^{2}}-d^{2} \; ln[\frac{\sqrt{1+d^{2}}+1}{\sqrt{1+d^{2}}-1}]),
\end{align}
while the second and fourth integrals are
 \begin{align}\label{Integ_ubt}
\int_{-\infty}^{\infty}\frac{1}{2}u_{b,t}^2\,dx
 =\int_{-\infty}^{\infty}\frac{1}{2}\dot{a}^2 e^{-2\kappa|x|}\,dx= \frac{1}{2\kappa}\dot{a}^2,\\
\int_{-\infty}^{\infty}\frac{1}{2}u_{b,x}^2\,dx
=\int_{-\infty}^{\infty}\frac{1}{2} (a(-\kappa) e^{-\kappa|x|}\; sgn(x))^2\,dx
= \frac{\kappa}{2}a^2.
\end{align}
The fifth integral is 
\begin{align} \int_{-\infty}^{\infty}(1-\gamma\delta(x))\psi(u_{k}+u_{b})\,dx
=\int_{-\infty}^{\infty}\psi(u_{k}+u_{b})\,dx-\gamma\psi(u_{k}(0,X)+a).\label{eq:int5}
\end{align}

Expanding $\psi(u_{k}+u_{b})$ via Taylor series around $a=0$, we get
\begin{multline}\label{NonlinPotExapnd}
\psi(u_{k}+u_{b})
\approx \psi(u_{k})-\frac{2(1-u_k)(\sqrt{(1-u_{k})^2+d^2}-\sqrt{1+d^2})}{\sqrt{(1-u_{k})^2+d^2}}u_{b}+\left(\frac{(1-u_{k})^2}{(1-u_{k})^2+d^2}\right.\\ 
\left.+\frac{d^2(\sqrt{(1-u_{k})^2+d^2}-\sqrt{1+d^2})}{((1-u_{k})^2+d^2)\sqrt{(1-u_{k})^2+d^2}}\right)u_{b}^{2}+O(a^3).
\end{multline}

Using (21,22,29), we write (\ref{NonlinPotExapnd}) as
\begin{multline}\label{NonlinPotentialExapnd}
\psi(u_{k}+u_{b})
\approx \psi(u_{k})-\underbrace{\frac{2\tanh\left(\frac{x-X}{\sqrt{2}C_{2}}\right)(\sqrt{\tanh^{2}\left(\frac{x-X}{\sqrt{2}C_{2}}\right)+d^2}-\sqrt{1+d^2})}{\sqrt{\tanh^{2}\left(\frac{x-X}{\sqrt{2}C_{2}}\right)+d^2}}a e^{-\kappa\;\left|x\right|}}_{T_1}+\\
\underbrace{\left(\frac{\tanh^2\left(\frac{x-X}{\sqrt{2}C_{2}}\right)}{\tanh^{2}\left(\frac{x-X}{\sqrt{2}C_{2}}\right)+d^2}+\frac{d^2\left(\sqrt{(\tanh^{2}\left(\frac{x-X}{\sqrt{2}C_{2}}\right))+d^2}-\sqrt{1+d^2}\right)}{(\tanh^{2}\left(\frac{x-X}{\sqrt{2}C_{2}}\right)+d^2)\sqrt{\tanh^{2}\left(\frac{x-X}{\sqrt{2}C_{2}}\right)+d^2}}\right)a^2 e^{-2\kappa\;\left|x\right|}}_{T_2}+O(a^3).
\end{multline}

The numerator of $T_1$ is a product of two terms. For $|x-X|\ll 1$, the first term is small while the second term remains bounded. For  $|x-X|\gg 1$, the second term is small while the first term remains bounded. Hence we assume $T_1\approx 0$.

Finally, $T_2$ is a sum of two terms. 
We only keep the contribution for the case when $|x-X|\gg 1$, and assume that the solitary wave-defect interaction is captured by the the $\delta$ function term in Eq. (\ref{eq:int5}). With these approximations, the fifth integral is $\int_{-\infty}^{\infty}\left[\psi(u_{k})+\left(\dfrac{e^{-2\kappa\;\left|x\right|}}{1+d^2}\right)a^{2}\right]\,dx-\gamma\psi(u_{k}(0,X)+a)$
\begin{align}=\frac{\sqrt{K_{r}}}{2\sqrt{2}}\left(2\sqrt{1+d^{2}}-d^{2} \; ln\left[\frac{\sqrt{1+d^{2}}+1}{\sqrt{1+d^{2}}-1}\right]\right)+ \frac{1}{(1+d^2)\kappa}a^{2}\nonumber-\gamma\psi\left(u_{k}(0,X)+a\right).
\end{align}

\section{Derivatives of $F$, $G$ and $R$}
The (partial) derivatives of $G(X)$, $F(X,a)$, and $R(a)$ are
\begin{equation}\label{dGdX}
\frac{dG}{dX}=\frac{-\sqrt{2}}{C_2}\sech^{2}\left({\frac{X}{\sqrt{2}C_2}}\right)\tanh\left({\frac{X}{\sqrt{2}C_2}}\right),
\end{equation}

\begin{equation}\label{dFdX}
\frac{\partial F}{\partial X}=\frac{\sqrt{2}}{C_2}\sech^{2}\left({\frac{X}{\sqrt{2}C_2}}\right)\left(a+2 \tanh\left({\frac{X}{\sqrt{2}C_2}}\right)-\frac{\sqrt{1+d^2} \left(\tanh\left({\frac{X}{\sqrt{2}C_2}}\right)+a\right)}{\sqrt{\left(\tanh\left({\frac{X}{\sqrt{2}C_2}}\right)+a\right)^2+d^2}}  \right),
\end{equation}

\begin{equation}\label{dFda}
\frac{\partial F}{\partial a}=2\left( \tanh\left({\frac{X}{\sqrt{2}C_2}}\right)-\frac{\sqrt{1+d^2} \left(\tanh\left({\frac{X}{\sqrt{2}C_2}}\right)+a\right)}{\sqrt{\left(\tanh\left({\frac{X}{\sqrt{2}C_2}}\right)+a\right)^2+d^2}}  \right),
\end{equation}

\begin{equation}\label{dRda}
\frac{dR}{da}=2a.
\end{equation}

\section{Fixed point analysis}\label{app:fixedpt}
In this section, we find conditions on parameters $C_0$ and $d$ such that the fixed point $(X^*=0,p_X^*=0,a^*=0,p_a^*=0)$ is always of type centre $\times$ center for all $0\leq \mu \leq 1$ and $0 \leq \gamma \leq 1$. This requires that $(X^*,a^*)$ should be a minimum of the potential energy 
\begin{equation}\label{Pertubed_Pot_Energy}
V(X,a)=\left(\frac{1}{(1+d^2)\kappa}+\frac{C_{0}^2\kappa}{2}\right)a^{2}-\gamma\left(R(a)+\mu F(X,a)+G(X)\right)+\frac{C_{0}A}{\sqrt{2}}.
\end{equation}
This condition is satisfied if the the Hessian of $V(X,a)$ is positive definite at $(X^*,a^*)$. This requires that both the eigenvalues of the Hessian are positive. The eigenvalues are given by
\begin{equation}\label{Eig_Hess}
\lambda_H=\frac{1}{2}\left[ (V_{,XX}+V_{,aa})\pm\sqrt{(V_{,XX}+V_{,aa})^2-4(V_{,XX}V_{,aa}-V_{,Xa}^2)}.  \right]
\end{equation}
The following three inequalities guarantee the positivity of both eigenvalues:
\begin{equation}\label{Hessian_P.d_Conditions}
V_{,XX}>0,V_{,aa}>0\text{, and } V_{,XX}V_{,aa}-V_{,Xa}^2>0,
\end{equation} 
where for the fixed point at the origin $(X^*=0,a^*=0):$
\begin{align}
V_{,XX}=\frac{\gamma}{C_{2}^2}\left[1+\mu \left(\frac{\sqrt{1+d^2}-2d}{d}\right)\right],\label{eq:Vxx}\\
V_{,aa}=2\gamma \left(\frac{\sqrt{1+d^2}}{d}\mu-1\right)+\left(\frac{2}{(1+d^2)\kappa}+\kappa\;C_{0}^2\right),\label{eq:Vaa}\\
\text{ and } V_{,Xa}=\sqrt{2}\frac{\gamma \mu}{C_2} \left(\frac{\sqrt{1+d^2}}{d}-1\right).\label{eq:VXa}
\end{align} 
To show that $V_{XX}>0$, we note that $\left(\dfrac{\sqrt{1+d^2}-2d}{d}\right)>-1$ for all $d>0$. Since $0\leq \mu\leq 1$, the result follows.

Next, we show that $V_{,aa}>0$ for $C_0>\gamma \sqrt{1-\frac{\omega_0^2}{4}}$. Squaring both sides, and using the relations $\omega_0^2=\frac{2}{1+d^2}$, and $\kappa=\dfrac{\gamma\omega_0^2}{2K_r}$, we get
\begin{align}
    C_0^2> \gamma^2-\gamma^2\frac{\omega_0^2}{4}\implies
    C_{0}^{2}\kappa+\frac{2}{(1+d^2)\kappa}-2\gamma\;\;>0
\end{align}
Since $0\leq \mu\leq 1$, the result follows.

Finally, one can verify by direct substitution that we can also ensure $V_{,XX}V_{,aa}-V_{,Xa}^2>0$ if we choose $C_0=\gamma m\sqrt{1-\frac{\omega_0^2}{4}}$, where
\begin{equation}\label{3rd_Cond}
    m>\max{\left(\sqrt{\frac{2(\mu-1)^2 - \frac{1}{1+d^2} \left[1+\left(\frac{\sqrt{1+d^2}}{d}-2 \right)\mu \right]}{2\left(1-2\mu+\frac{\sqrt{1+d^2}}{d}\mu\right)\left(1-\frac{1}{2(1+d^2)}\right)}}\;,1\right)}.
\end{equation}



\section{Omitted details of Melnikov analysis}
To finish the Melnikov analysis, and confirm the existence of heteroclinic tangles in the system, we need to further prove that $\dfrac{dM(\theta_0)}{d\theta_0}|_{\theta_0=\pi/2}\neq 0$, where \begin{equation}\label{eq:Melfuntheta0}
M(\theta_0)= \frac{2\gamma}{C_2}\sqrt{\frac{\sqrt{2}\gamma \; C_0}{A}} \int_{-\infty}^{\infty} Q(t) [M_1(t,\theta_0)+M_2(t)-M_3(t,\theta_0)] dt,   
\end{equation}
 $Q(t)= \dfrac{1}{(1+N^2 t^2)\sqrt{(1+N^2 t^2)}},
M_1(t,\theta_0)=S \sqrt{\omega I^0} cos(t+\theta_0), 
M_2(t)=2 \dfrac{N t}{\sqrt{(1+N^2 t^2)}},$ and \\$
M_3(t,\theta_0)=\dfrac{\sqrt{1+d^2}\left(S \sqrt{\omega I^0} cos(t+\theta_0)+ \dfrac{N t}{\sqrt{(1+N^2 t^2)}}\right)}{\sqrt{d^2+\left(S \sqrt{\omega I^0} cos(t+\theta_0)+ \dfrac{N t}{\sqrt{(1+N^2 t^2)}}\right)^2}}.$  

\begin{figure}
    \centering
    \includegraphics[width=.45\textwidth]{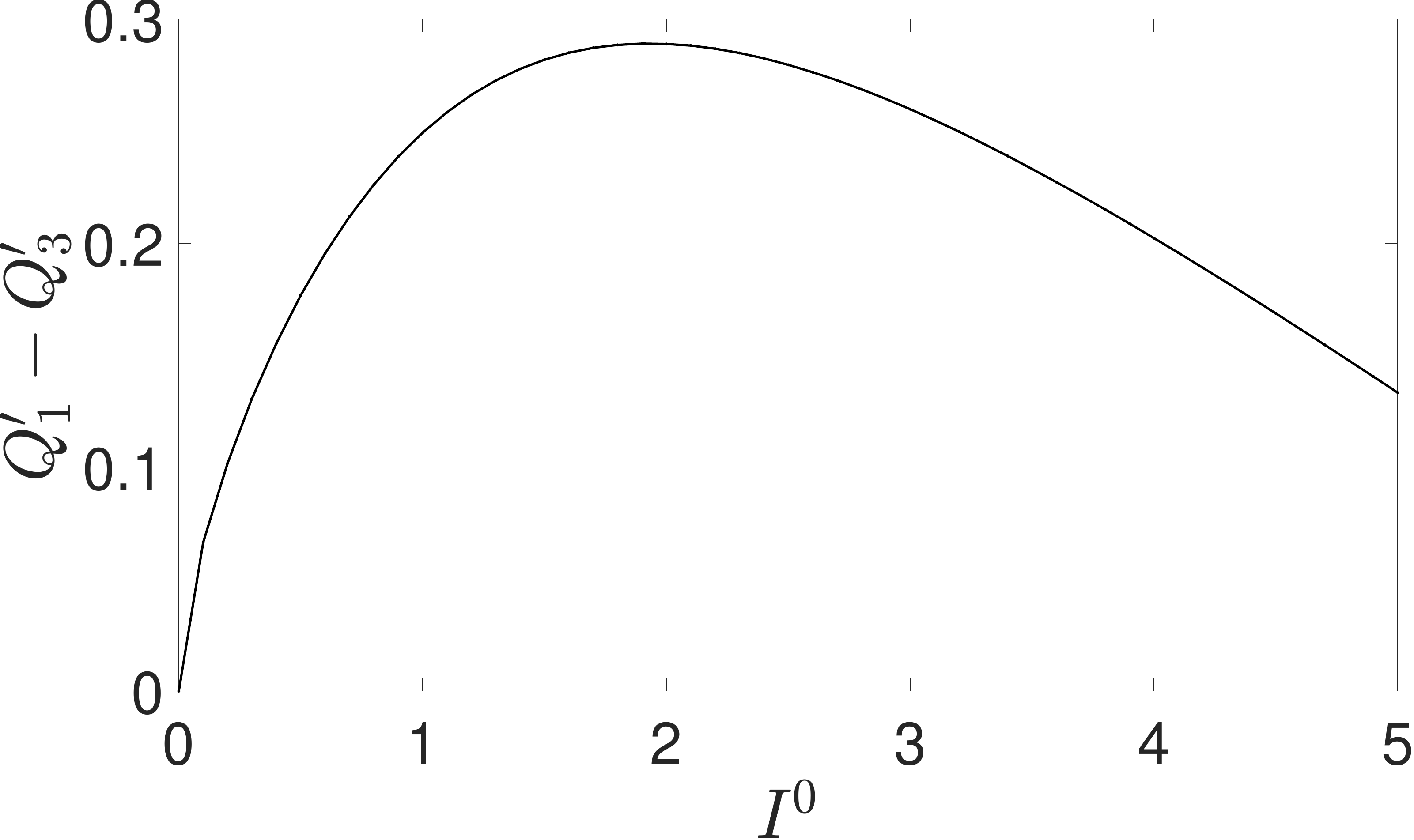}
    \caption{\footnotesize $\dfrac{dM(\theta_0)}{d\theta_0}|_{\theta_0=\pi/2}$ as a function of the size of the periodic orbit in the unperturbed system. }
    \label{fig:Melnikov_Derivative}
\end{figure}

From above, we obtain $\dfrac{dM(\theta_0)}{d\theta_0}|_{\theta_0=\pi/2}=Q_1^{'} - Q_3^{'}$, where $Q_1^{'}=-\int_{-\infty}^{\infty}\dfrac{S\sqrt{\omega I^0}}{(1+N^2 t^2)\sqrt{(1+N^2 t^2)}} \cos(t)dt$, and 
    \begin{multline}
    Q_3^{'}=\int_{-\infty}^{\infty}
    \dfrac{\sqrt{1+d^2}S\sqrt{\omega I^0}\cos{t}}{(1+N^2t^2)\sqrt{(1+N^2t^2)}\sqrt{d^2+\left(\dfrac{n t}{\sqrt{1+n^2 t^2}}-S\sqrt{\omega I^0}\sin{t}\right)^2}}\times\\
    \left(\dfrac{\left(\dfrac{Nt}{\sqrt{1+N^2 t^2}}-S\sqrt{\omega I^0}\sin{t}\right)^2}{d^2+\left(\dfrac{N t}{\sqrt{1+N^2 t^2}}-S\sqrt{\omega I^0}\sin{t}\right)^2}-1\right)dt.
    \end{multline}
$Q_1'$ can be analytically computed to yield
\begin{align} Q_1^{'}=-2\dfrac{S\sqrt{\omega I^0}K_1(\dfrac{1}{N})}{N^2},\end{align}
where $K_1$ is the modified Bessel function of the second kind. Since the second integral could not be computed analytically, we show the numerical results in Fig. \ref{fig:Melnikov_Derivative}. This computation confirms that $\dfrac{dM(\theta_0)}{d\theta_0}|_{\theta_0=\pi/2}> 0$ for the parameters relevant to this study.   
\clearpage
\bibliography{refs}
\end{document}